\documentclass[12pt,a4paper]{article}
\usepackage{ifpdf}
\usepackage{times}
\usepackage[scale={0.8,0.9},centering,includeheadfoot]{geometry}
\usepackage{enumitem}
\usepackage{subcaption}
\usepackage{multicol}
\usepackage[dvipsnames]{xcolor}
\definecolor{darkblue}{RGB}{24, 45, 158}
\usepackage{lscape}
\usepackage{booktabs}
\usepackage[colorlinks=true,citecolor=darkblue,linkcolor=blue,urlcolor=darkblue]{hyperref}
\usepackage{float}
\usepackage{pgf}

\usepackage{amsmath,amssymb,xspace,amsthm,bm}
\usepackage{hyperref}

\usepackage[colon,longnamesfirst]{natbib}

\newtheorem{definition}{Definition}

\newtheorem{example}{Example}

\newtheorem{model}{Model}
\newtheorem{remark}{Remark}
\newtheorem{proposition}{Proposition}

\newcommand*\diff{\mathop{}\!\mathrm{d}}

\def\qedbox{\ifvmode\else\unskip\fi~\penalty10000
    \hfill{\large$\blacksquare$}}

\begin{document}
\bibliographystyle{apalike}

\title{\vspace{-15mm}\fontsize{24pt}{10pt}\selectfont\textbf{A Standardization Procedure to Incorporate Variance Partitioning Based Priors in Latent Gaussian Models}}

\author{
    Luisa Ferrari \\
    Department of Statistical Sciences\\    
    University of Bologna\\
    \and
    Massimo Ventrucci \\
	Department of Statistical Sciences\\    
    University of Bologna\\
}
    
\maketitle

\begin{abstract}
Latent Gaussian Models (LGMs) are a subset of Bayesian Hierarchical models where Gaussian priors, conditional on variance parameters, are assigned to all effects in the model. LGMs are employed in many fields for their flexibility and computational efficiency. However, practitioners find prior elicitation on the variance parameters challenging because of a lack of intuitive interpretation for them. Recently, several papers have tackled this issue by rethinking the model in terms of variance partitioning (VP) and assigning priors to parameters reflecting the relative contribution of each effect to the total variance. So far, the class of priors based on VP has been mainly deployed for random effects and fixed effects separately. This work presents a novel standardization procedure that expands the applicability of VP priors to a broader class of LGMs, including both fixed and random effects. 
We describe the steps required for standardization through various examples, with a particular focus on the popular class of intrinsic Gaussian Markov random fields (IGMRFs). The practical advantages of standardization are demonstrated with simulated data and a real dataset on survival analysis.
\end{abstract}


\textbf{Email for Correspondence}: \href{mailto:luisa.ferrari5@unibo.it}{luisa.ferrari5@unibo.it}

\section{Introduction}\label{sec:intro}

Latent Gaussian Models (LGMs, \citet{RM}) are a popular class of Bayesian Hierarchical models used by applied scientists in many fields for their flexibility and ability to include linear and smooth effects of continuous covariates, spatial and temporal random effects and more complex structures such as interactions. The linear predictor for an observation $i$ is structured as:
\begin{equation*}
       \eta_i =\mu+ \sum_{p=1}^P x_{ip} \beta_p + \sum_{r=1}^Rf_r(z_{ir})
\end{equation*}
where $\beta_p\sim N(0,\sigma_p^2), p= 1, \ldots , P$  are referred to as fixed effects, associated to $P$ covariates, and $\boldsymbol{f_r}\sim N(\boldsymbol{0}, \boldsymbol{\Sigma}_r), r=1,\ldots\,R$, are called random effects. Both fixed and random effects in an LGMs are Gaussian conditional on the parameters in their covariance. Variances for fixed effects, $\sigma_p^2$, are typically fixed at large values. For random effects, $\boldsymbol{\Sigma}_r$ is often expressed as a correlation matrix scaled by a random variance $\sigma_r^2$ who is assigned a prior (hence $\sigma_r^2$ is an hyper-parameter). Along with proper Gaussian processes, LGMs frequently make use of Intrinsic Gaussian Markov Random Fields (IGMRFs, \citet{RH} and \citet{FK04}). IGMRFs, which include as special cases the random walk of order 1 (RW1) and 2 (RW2) and the intrinsic conditional autoregressive process (ICAR, \citet{besag1995conditional}), are  especially employed as priors for spatial and temporal random effects. 

Prior specification on the variances of the random effects, $\sigma_r^2$, is a long discussed problem due to the poor interpretability of these parameters, especially for IGMRFs (\citet{SR14}).
Independent Inverse Gamma priors have been found to lead to overfitting (\citet{G06}, \citet{lunn2009rejoinder}, \citet{S17}) and are now being replaced by Penalized Complexity (PC) priors (\cite{S17}) in the \texttt{R} package \texttt{INLA} (\citet{RM}), which provides fast approximate inference for LGMs.

While variances are hard to elicit, experts often have some intuition about the relative importance of the different effects in terms of variance contribution. Recently, multiple works have tried to leverage this underlying information by designing joint priors on the variance parameters, either the $\sigma_p^2$'s or the $\sigma_r^2$'s above mentioned. Specifically, a joint prior can be specified using a reparameterization of the original parameters into a total variance and a set of proportions of variance. Proportions are more intuitive parameters for users, as they directly indicate the relative importance of effects on a (0-1) scale, thereby facilitating prior specification. This novel approach can be referred to generally as \textit{Variance Partitioning  (VP)}, and in this paper the term \textit{VP priors} is used to define the class of priors derived using this approach.

Two main lines of research can be recognized in the literature on VP priors. The first one follows the work by \citet{F20} who introduce the class of Hierarchical Decomposition (HD) priors, which can be seen as a subclass of VP priors focusing exclusively on random effects (i.e. variance partitioning is applied to $\sigma_r^2$'s only). HD priors can be applied to all elements in the linear predictor or in a hierarchical manner through consecutive splits of the linear predictor; for applications see \citet{hem2021robust} and \citet{FV22}. Similarly, in disease mapping \citet{wakefield2007disease} and \citet{riebler2016intuitive} have proposed priors on the variance contribution of spatially structured and unstructured components (\citet{BYM}).The second line of research based on \citet{R2D2} considers priors on a measure of goodness-of-fit $R^2$, denoted as R2D2 priors, which initially focused exclusively on linear effects (i.e. variance partitioning is applied to $\sigma_p^2$'s only) but have recently been extended to mixed model cases (\citet{R2D2M2}, \citet{R2D2_GLMM}, \cite{R2D2_space}).
Similarly,  \citet{bhattacharya2015dirichlet} define a global-local shrinkage prior for sparse regression. 

It is worth noting that the applicability of VP priors in LGMs relies on the user being able to interpret the variance parameters as \emph{variance contributions} of the corresponding effects. Thus, all effects in the model should somehow live in the same scale so that, for instance, $\sigma_r=1, \forall r$, means equal a priori contribution for all random effects. In other words, an \emph{intuitive interpretation} of the variance of each effect is required to safely use VP priors. 
Many popular effects (called heterogeneous by \citet{F20}) do not satisfy such interpretation requirement. Notable examples are some popular IGMRFs (e.g., ICAR, RW1, RW2); as noted by \citet{SR14}, the variance of an IGMRF strongly depends on the domain in which the process is observed. \citet{SR14} proposed a solution based on rescaling the strucure matrix, which is recommended by \citet{F20} for inclusion of IGMRFs in the HD framework. R2D2 recommends standardization of all covariates, which solves the issue in an LGM with only linear effects. It is not clear how to guarantee the interpretation requirement in a broader LGM framework including both fixed and random effects.

In this paper, we aim to extend the applicability of VP priors to a class of Latent Gaussian Models including both fixed and random effects. We start by defining variance contribution for fixed and random effects separately and then derive a novel \textit{standardization procedure} that guarantees an exact match between variance parameters and variance contributions. We prove analytically that the proportions of variance parameters are correctly interpreted after standardization (Remark \ref{remark:VP}). The benefits of the proposed standardization are confirmed by simulations. 
On the practical side, our primary interest is to enable the use of VP priors in realistic settings. For this reason, we illustrate the proposed standardization through various examples frequently encountered in LGMs, such as linear and smooth effects of continuous covariates, spatial and temporal random effects, etc. We put emphasis on IGMRFs as these are popularly used not only to model spatial and temporal dependence but also to construct smooth effects via P-splines (\citet{EM}). We argue that a preliminary adjustment is necessary for P-splines before standardization. We check via simulation the usefulness of such adjustment both in the context of VP priors and PC priors. 
For a user-friendly practical implementation of standardization, the \texttt{scaleGMRF} R package has been developed and made publicly available at \url{https://github.com/LFerrariIt/scaleGMRF}. 

The remainder of the paper is organized as follows. Section \ref{sec:basics} sets the background by defining the class of LGMs under study, VP priors and the different notions of variance contribution underlying HD priors and R2D2 priors. In Section \ref{sec:standardization}, we propose our definition of the concept of intuitive interpretation for variance parameters and present the standardization procedure. In Section \ref{sec:IGMRFs}, we describe the challenges behind the standardization of IGMRFs focusing on the P-spline case. Section \ref{sec:results} reports two simulation studies and an application to real data. The paper closes with a discussion in Section \ref{sec:discussion}.

\section{Background}\label{sec:basics}

\subsection{Latent Gaussian Models}\label{sec:LGMs}
In this work, we consider a popular subclass of LGMs in which the response is linked to the latent parameters with a univariate link function (\citet{BHM_ref}), 
and each effect covariance matrix $\boldsymbol{\Sigma}$ is expressed as $\sigma^2 \boldsymbol{Q}^*$, where $\sigma^2$ is a random scalar and $\boldsymbol{Q}$ is a known \emph{structure} matrix (i.e., the inverse covariance matrix given $\sigma^2=1$) which can take various forms according to the different type of GMRF or IGMRF considered.
 
\begin{model}[LGMs]\label{def:model} 
Let $\boldsymbol{X}=[X_1,\ldots ,X_J]$ be a set of covariates with $\boldsymbol{X}\sim \pi(\boldsymbol{x})$, and possible realizations $x\in \mathcal{X}_j$ for $j=1,\ldots ,J$. Let a response $Y\sim \text{Dist}(\eta,\boldsymbol{\psi})$ have an exponential family distribution, where $\eta$ corresponds to a transformation of the expected value of $Y$ given $\boldsymbol{\psi}$ and:
\begin{align*}
       \eta &=\mu+\sum_{j=1}^Jf_j(X_j).
\end{align*}
Each effect $f_j(X_j)$ for $j=1,\ldots ,J$ is defined as: 
\begin{align*}
f_j(X_j)&=\boldsymbol{D}^T_j(X_j)\boldsymbol{u}_j; \ \ \ \
     \boldsymbol{u}_j|\sigma^2_j   \sim N_{K_j}(\boldsymbol{0},\sigma^2_j\boldsymbol{Q}_j^*)
\end{align*}
where the basis $\boldsymbol{D}_j(X_j)$ is a column vector containing $K_j$ known functions evaluated at $X_j$, the structure matrix $\boldsymbol{Q}_j$ is known, and the scalar $\sigma^2_j$ is the variance parameter. The parameters of the model that require prior specification are $\boldsymbol{\sigma}=[\sigma^2_1,\ldots ,\sigma^2_J]$, along with $\mu$ and $\boldsymbol{\psi}$.
\end{model}
$\boldsymbol{Q}_j^*$ represents the generalized inverse of $\boldsymbol{Q}_j$, so that Model \ref{def:model} can include improper Gaussian effects, as long as the appropriate constraints are imposed (see Section \ref{sec:IGMRFs}). Linear effects can be specified in Model \ref{def:model} setting $K_j=1$ and the basis to a single identity function, i.e. $D_j(X_j)=X_j$. If some $\boldsymbol{Q}_j$ are not actually fixed but controlled by unknown parameters (e.g. the range in the Matern process), they must be assumed fixed to a reasonable value for the purpose of prior specification on $\boldsymbol{\sigma}$ (\citet{F20}). Finally, note that the decision to explicitly consider the covariates 
$\boldsymbol{X}$ as random variables is indispensable to the development of the theory in Section \ref{sec:standardization}.

\subsection{Variance partitioning priors}\label{sec:VP_priors}
The traditional approach to prior specification on $\boldsymbol{\sigma}$ sets the variance parameters of linear effects to large \emph{fixed} values, to ensure flat priors on their coefficients, while the variance parameters of the remaining effects are usually treated as \emph{random} and assigned independent and identical priors. 

Alternatively, available prior information could be better exploited by treating all variance parameters as random, and specifying a joint prior on $\boldsymbol{\sigma}$ that reflects assumptions about the relationships between the contributions of the different effects. Specifying such prior is much easier using a convenient transformation of the variance parameters into a single total variance and a set of proportions of this total variance: this reparametrization can be called \emph{variance partitioning}, and priors specified using this method can be referred to as VP priors.
\begin{definition}[VP priors]\label{def:VP}
Consider Model \ref{def:model}. We define the VP parameters as: 
\begin{equation}
\label{eq:VP_params}
\begin{gathered}
    V=\sum_{j=1}^J\sigma^2_j,\\
\boldsymbol{\omega}=\left[\omega_1=\dfrac{\sigma^2_1}{V},\ldots ,\omega_{J-1}=\dfrac{\sigma^2_{J-1}}{V},\omega_J=1-\sum_{j=1}^{J-1}\omega_j\right].
\end{gathered}
\end{equation}
We call VP those priors implied on the original variance parameters by the specification of independent priors on the VP parameters, i.e.:
\begin{equation*}
\pi(\sigma^2_1,\ldots ,\sigma^2_J)=\pi(V)\pi(\boldsymbol{\omega})|\boldsymbol{J}|.
\end{equation*}
where $\boldsymbol{J}$ represents the Jacobian associated with the transformation $\boldsymbol{\sigma}\rightarrow V,\boldsymbol{\omega}$.
\end{definition}
As defined, the class of VP priors is extremely broad, and can cover most of the works cited in Section \ref{sec:intro}. The two main lines of research using this approach, i.e. HD and R2D2 priors, have so far specified VP priors in very different ways. Considering first the work by \citet{F20}, the HD approach has proposed the use of a Jeffreys or a PC prior on $V$ (according to the likelihood), and a more complex prior on $\boldsymbol{\omega}$. 
Specifically, $\boldsymbol{\omega}$ is further transformed in separate sets of proportions, obtained through a case-specific hierarchical decomposition of the variance via subsequent splits. The joint prior on $\boldsymbol{\omega}$ is then obtained via independent priors on the sets of proportions generated at each split: the authors suggest the use of either Dirichlet or PC priors on the vectors of proportions to reflect different kinds of prior beliefs. 
The $R2D2$ approach has proposed instead the use of a simple symmetric Dirichlet prior on $\boldsymbol{\omega}$. On the other hand, the prior on $V$ is not directly specified, but rather implied imposing a Beta prior on the measure of goodness-of-fit $R^2$ (\citet{R2D2_GLMM}), defined as a function of $V$ and the additional likelihood parameters $\boldsymbol{\psi}$ from Model \ref{def:model}.

\subsection{The notion of variance contribution in HD and R2D2 priors}\label{sec:int_int}

The appeal of VP priors lies in the possibility of obtaining parameters that are more easily interpretable for the user when compared to the original ones. However, $V$ and $\boldsymbol{\omega}$ are not always equal to their \textit{intuitive interpretation}, i.e. respectively as the total variance due to the $J$ effects and the set of proportional contributions of the individual effects to this total variance. Such intuitive interpretation of $V$ and $\boldsymbol{\omega}$ is only guaranteed if each $\sigma^2_j$ actually matches its own intuitive interpretation, implicitly defined by the user as the variance contribution of its corresponding effect.

The literature on VP priors has so far defined the concept of variance contribution of an effect in different ways, either directly or implicitly. 
Considering the work of \citet{R2D2_GLMM} as the main reference, the R2D2 approach has so far only worked with models where $\sigma^2_j=\text{E}_{X_j}\{Var_{\boldsymbol{u}_j}[f_j(X_j)|X_j,\sigma^2_j]\}$ for all effects. Hence, the authors have implicitly defined the intuitive interpretation of $\sigma^2_j$ using what can be called the \emph{marginal variance} of the effect, i.e. the variance of $f(X_j)=\boldsymbol{D}^T_j(X_j)\boldsymbol{u}_j$ computed by marginalizing out both $X_j$ and $\boldsymbol{u}_j$ and viewed as a function of $\sigma^2$ ($Var_{X_j,\boldsymbol{u}_j}[f_j(X_j)|\sigma^2_j]$).
On the other hand, the HD framework explicitly requires that each variance parameter must be (at least approximately) equal to the \emph{conditional variance} of its corresponding effect given $X$, i.e. $Var_{\boldsymbol{u}_j}[f_j(X_j)|\sigma^2_j,X_j=x]$, although again, many popular effects do not satisfy this requirement either. Instead of limiting the applicability of HD priors, \citet{F20} employed the scaling procedure proposed by \citet{SR14}, which ensures that each $\sigma^2_j$ becomes equal to the geometric mean (GM) of the conditional variance. This approach effectively corresponds to defining the variance contribution of an effect as $\text{GM}_{X_j}\{Var_{\boldsymbol{u}_j}[f_j(X_j)|X_j]\}$. However, the geometric mean scaling solution, initially developed for IGMRFs, exhibits some important limitations. Notably, it fails for linear effects, which therefore remain outside the scope of application of HD priors, along with other popular effects designed for continuous covariates.
The lack of a clear and consistent definition of variance contribution, coupled with the absence of a general and reliable method to align variance parameters with their intended interpretation, has severely limited the widespread adoption of VP priors, leading to a fragmented and inconsistent application of the method.

\section{Standardization procedure}\label{sec:standardization}
In this section, we establish a formal definition of variance contribution and introduce a standardization procedure to ensure alignment between variance parameters and their intuitive interpretations. Our proposal critically relies upon a modern rethinking of the traditional concepts of fixed and random effects.

\subsection{Redefinition of fixed and random effects}
Although all effects in a Bayesian LGM are random, it is common practice to label the effects of Model \ref{def:model} as fixed (i.e. linear) or random (i.e. other effects), according to the treatment of the variance parameters as either fixed or random. 
The VP approach considers all variance parameters as random. Hence, the traditional criterion for the fixed/random classification does not apply.
Nevertheless, the literature recognizes the existence of a deeper intrinsic binary categorization of model effects. For instance, \citet{BDA} state in Section 15.6 that:

\begin{quotation}
[\ldots ] much of the statistical literature on fixed and random effects can be fruitfully re-expressed in terms of finite-population and super-population inferences [\ldots ]. The difference between fixed and random effects is thus not a difference in inference or computation but in the ways that these inferences will be used.
\end{quotation} 
Similarly, \citet{H13} proposes to distinguish random effects (i.e. having a random variance parameter) in those whose levels  \emph{``[\ldots] are draws from a population, and the draws are not of interest in themselves [\ldots]"}, and those whose \emph{``[\ldots] levels themselves are of interest".}
Adopting this modern perspective, we propose a redefinition of the fixed/random effects classification on the basis of the concept of \emph{inferential interest}.
\begin{definition}[Fixed and random effects]\label{def:fixed_random}
Consider each effect $f_j(X_j)$ from Model \ref{def:model}. If inference about $\boldsymbol{u}_j$ (or transformations) is of interest, the $j$ effect shall be referred to as \emph{fixed}; if instead, inference about the parameter $\sigma^2_j$ (or transformations) is of interest, the $j$ effect shall be referred to as \emph{random}. Model \ref{def:model} shall be specified such that $\boldsymbol{\theta}=[\boldsymbol{u}_1,\ldots ,\boldsymbol{u}_L,\sigma^2_{L+1},\ldots ,\sigma^2_J]$ contains all the parameters of interest for the user, i.e. the first $L$ effects will be fixed, and the last $J-L$ will be random.
\end{definition}
Under Definition \ref{def:fixed_random}, linear effects are clearly considered fixed, as the inferential focus is always on the slope coefficients. Other effects, such as non-linear, spatial or temporal ones, will be considered fixed or random according to the researcher's specific inferential objectives. Hereafter, we use the terms fixed and random according to Definition \ref{def:fixed_random}.

\subsection[section3.2]{Intuitive interpretation of $\boldsymbol{\sigma}$}

We argue that the user has a different definition of variance contribution for fixed and random effects. For example, the variance contribution of linear effects (always considered fixed) is usually estimated conditioning on the linear coefficient itself to obtain what is often called the \textit{explained variance} $Var_X[X\beta]$. On the other hand, the amount of variance due to group i.i.d. effect (often considered as random) is usually estimated directly through the variance parameter $\sigma^2$. More generally, we adopt the terminology of \citet{BDA} to argue that the user intends variance contribution as the:
\begin{itemize}
    \item \textit{finite-population variance} for fixed effects, i.e. $Var_{X_j}[f_j(X_j)|\boldsymbol{u}_j]$;
    \item \textit{super-population variance} for random effects, i.e. $Var_{X_j,\boldsymbol{u}_j}[f_j(X_j)|\sigma^2_j]$.
\end{itemize}
This distinction can be neatly summarized using the concept of \textit{variance of interest}, defined as the variance of the effect conditional on the parameters of interest, i.e. $Var_{X_j,\boldsymbol{u}_j}[f_j(X_j)|\boldsymbol{\theta}]$ with $\boldsymbol{\theta}$ as in Definition \ref{def:fixed_random}. However, the variance of interest cannot be directly used as the basis for the definition of the intuitive interpretation of $\sigma_j$, as it is not a function of $\sigma^2_j$ for fixed effects. Hence, we propose to use instead the \textit{expected variance of interest}, defined as: $E_{\boldsymbol{\theta}}\{Var_{X_j,\boldsymbol{u}_j}[f_j(X_j)|\boldsymbol{\theta}]|\boldsymbol{\sigma}\}$.
Deriving the expected variance of interest separately for fixed and random effects, we can formally define the intuitive interpretation of the $\sigma^2_j$ parameters for all effects of Model \ref{def:model}.

\begin{definition}[Intuitive interpretation of $\boldsymbol{\sigma}$ parameters]\label{def:int_int}\ \\
Consider Model \ref{def:model} with parameters of interest $\boldsymbol{\theta}=[\boldsymbol{u}_1,\ldots ,\boldsymbol{u}_L,\sigma^2_{L+1},\ldots ,\sigma^2_J]$. We say that $\sigma^2_1,\ldots ,\sigma^2_J$ match their intuitive interpretation if the following conditions hold:
\begin{itemize}
    \item for fixed effects $(j=1,\ldots ,L)$
    \begin{gather}\label{eq:fe_int_req} 
         \sigma^2_j=E_{\boldsymbol{u}_j}\{Var_{X_j}[f_j(X_j)|\boldsymbol{u}_j]|\sigma^2_j\};
    \end{gather}
     \item for random effects $(j=L+1,\ldots ,J)$
    \begin{gather}\label{eq:re_int_req}      
    \sigma^2_j= Var_{X_j,\boldsymbol{u}_j}[f_j(X_j)|\sigma^2_j].
    \end{gather} 
\end{itemize}
\end{definition}
Definition \ref{def:int_int} is consistent with the suggestion in \citet{F20} of considering the explained variance for the introduction of fixed effects in the HD framework, since Eq. (\ref{eq:fe_int_req}) corresponds to the \textit{expected explained variance}.  On the other hand, Definition \ref{def:int_int} is also coherent with the R2D2 approach since Eq. (\ref{eq:re_int_req}) can be proven to be equivalent to the implicit definition of variance contribution in the R2D2 literature. 

Note that Definition \ref{def:int_int} relies on the probability assumption made in Model \ref{def:model} about the covariates $\boldsymbol{X}$, which therefore directly affects the definition of intuitive interpretation of the $\sigma^2_j$ parameters. In this context, $\pi(\boldsymbol{x})$ does not need to be specified as an estimate for the actual probability distribution of the covariates in the phenomenon, such as for example using the empirical distribution; instead, it should reflect the user's intuition about the variance contribution of the covariates. A sensible default approach may consist in the use of Uniform distributions over given supports of interest for each covariate; 
in fact, this option also proves to be convenient for various popular effects (see Examples \ref{ex:2}-\ref{ex:4}).

\subsection{Centering and scaling}

Inspired by the work of \citet{SR14}, we propose a standardization procedure that ensures that the conditions of Definition \ref{def:int_int} are satisfied for all effects of Model \ref{def:model}.

\begin{proposition}[Standardization procedure]\label{prop:standardization}
Consider Model \ref{def:model} with parameters of interest $\boldsymbol{\theta}=[\boldsymbol{u}_1,\ldots ,\boldsymbol{u}_L,\sigma^2_{L+1},\ldots ,{\sigma^2_J}]$. Definition \ref{def:int_int} is satisfied after the implementation of the following two steps.
\begin{enumerate}
\item \textbf{0-mean constraint on fixed effects}. Each fixed effect $j=1,\ldots ,L$ is constrained such that:
\begin{equation*}
\begin{aligned}
 E_{X_j}[f_j(X_j)|\boldsymbol{u}_j]=0.
\end{aligned}
\end{equation*}
\item \textbf{Scaling}. Each effect $j=1,\ldots ,J$ is replaced by $\widetilde{f}_j(X_j)=\dfrac{f_j(X_j)}{\sqrt{C_j}}$ where: 
\begin{equation}
\label{eq:scaling_constant}
\begin{aligned}
  C_j&= Var_{X_j,\boldsymbol{u}_j}[f_j(X_j)|\sigma^2_j=1]\\
  &=E_{X_j}\left[\boldsymbol{D}^T_j(X_j)\boldsymbol{Q}_j^*\boldsymbol{D}_j(X_j)\right].
\end{aligned}
\end{equation}
\end{enumerate} 
See the proof in Section A of the Supplementary Material.
\end{proposition}

In other words, Proposition \ref{prop:standardization} states that scaling by the appropriate constant $C_j$ is sufficient for random effects, as this guarantees the equality between $\sigma^2_j$ and the super-population variance (Eq. (\ref{eq:re_int_req})). On the other hand, both steps of Proposition \ref{prop:standardization} are necessary for fixed effects to satisfy Eq. (\ref{eq:fe_int_req}).
Proposition \ref{prop:standardization} enables us to derive interpretable expressions for the VP parameters, as detailed in the following remark.

\begin{remark}[Interpretation of the VP parameters]\label{remark:VP}\ \\
Consider Model \ref{def:model} with parameters of interest $\boldsymbol{\theta}=[\boldsymbol{u}_1,\ldots ,\boldsymbol{u}_L,\sigma^2_{L+1},\ldots ,{\sigma^2_J}]$. If the conditions from Definition \ref{def:int_int} hold, then the VP parameters from Eq. (\ref{eq:VP_params}) are equal to:
\begin{align*}
   V&= E_{\boldsymbol{\theta}}\{Var_{\boldsymbol{X},\boldsymbol{u}_1,\ldots ,\boldsymbol{u}_J}[\eta|\mu,\boldsymbol{\theta}]|\boldsymbol{\sigma}\},
   \\
   \omega_j&= \dfrac{E_{\boldsymbol{\theta}}\{Var_{X_j,\boldsymbol{u}_j}[f_j(X_j)|\boldsymbol{\theta}]|\boldsymbol{\sigma}\}}{V}  \quad j=1,\ldots ,J.
\end{align*}
See the proof in Section B of the Supplementary Material.
\end{remark}

Remark \ref{remark:VP} proves how the parameter $V$ can be correctly interpreted after standardization as the expected variance of interest of the linear predictor, while the entries of $\boldsymbol{\omega}$ correctly represent the proportional contributions of each effect to $V$. The procedure is called \textit{standardization} as it can be proven that classical standardization is a special case that arises for linear effects.

\begin{example}[Linear effects]\label{ex:1}
Consider an effect $f(X)=X u$ for a covariate $X$, where $
    u|\sigma^2\sim N(0,\sigma^2)$. Since linear effects are always considered fixed, Proposition \ref{prop:standardization}  requires the application of both steps of the standardization procedure. The 0-mean constraint can be simply imposed redefining the single basis function as $X-E[X]$, while the scaling constant is found to be $C=Var_X[X]$. As a consequence, a linear effect after the application of Proposition \ref{prop:standardization} becomes: 
\begin{gather*}
 \widetilde{f}(X)= \dfrac{X-E_X[X]}{\sqrt{Var_X[X]}} u
\end{gather*}
which corresponds to a linear effect on the standardized version of covariate $X$. This result shows the coherence of Proposition \ref{prop:standardization}  with the approach taken in the R2D2 literature. 
\end{example}
Linear effects represent a special case in which the standardization is straightforward. The correct application to more complex effects requires special care in considering some technical aspects affecting both steps of the procedure.

\subsubsection*{0-mean constraint}
If the process has a single basis function $D(X)$ as in Example \ref{ex:1}, the 0-mean constraint can be imposed simply redefining it as $D(X)-E_X[D(X)]$. For a general process with more than one basis function, the 0-mean constraint can be imposed by deriving the distribution of $\boldsymbol{u}$ conditional on the linear constraint $\boldsymbol{a}^T\boldsymbol{u}=0$, where $\boldsymbol{a} = E_{X}[\boldsymbol{D}(X)]$.

\begin{example}[Random intercepts]\label{ex:2}
Consider a categorical $X$ with K levels and $\pi(X=k)=p_k$. An effect for $X$ can be expressed using $K$ independent and identically distributed coefficients $u_1,\ldots ,u_K$, each of them being linked to a level of the covariate through basis functions $D_k(X)=\mathbb{I}[X=k]$:
 \begin{gather*}\label{eq:random_intercept}
     f(X)=\sum_{k=1}^K \mathbb{I}(X=k) u_k; \ \ \ \
     \boldsymbol{u}|\sigma^2 \sim N(\boldsymbol{0},\sigma^2\boldsymbol{I}).
 \end{gather*}
If the effect is considered random, the standardization procedure has no impact as $C=1$. However, there might be cases where $u_1,\ldots ,u_K$ are directly of interest, i.e. the effect should be treated as a fixed one. In this second scenario, the 0-mean constraint is such that $\boldsymbol{a}=[p_1,\ldots ,p_K]^T$ and the scaling constant is found to be: $C=1-\frac{\sum_{k=1}^K p_k^3}{\sum_{k=1}^K p_k^2}$.
If $X$ is Uniformly distributed with $p_k=1/K,\;\forall k$, then $C=(K-1){K}^{-1}$. Hence, $C$ converges to 1 as the number of levels grows, i.e. as the variance in the mean of the process goes to $0$ and the the 0-mean constraint becomes less and less relevant. 
See the proofs for all results in Section C of Supplementary Material.\end{example}

\subsubsection*{Scaling constant}
The scaling step of Proposition \ref{prop:standardization} can be implemented in practice either dividing the basis matrix by $\sqrt{C_j}$ or by multiplying the structure matrix by $C_j$. The computation of the scaling constants $C_j$ requires taking into account different aspects. First, all linear constraints imposed on the effect (including the ones from the 0-mean constraint step) must be considered before the application of the scaling procedure (see Section \ref{sec:IGMRFs}). Secondly, there is no guarantee that $C_j$ will be non-null and finite for all potential models, as it is a function of $\pi(x_j),\boldsymbol{D}_j(X_j),\boldsymbol{Q}_j$: hence, the couples $f_j(X_j)$ and $\pi_j(x)$ must be designed to ensure that $0<C_j<\infty,\;\forall j$. Thirdly, it might be easier to approximate the values of $C_j$ using a Monte Carlo simulation, sampling $N$ values $x_1,\ldots ,x_N$ from $X_j\sim \pi_j(x)$ to compute an estimate $
    \widehat{C}_j=N^{-1}\sum_{i=1}^N \boldsymbol{D}^T_j(x_i)\boldsymbol{Q}^*_j\boldsymbol{D}_j(x_i)$. 
    
Noting that $C_j=E_{X_j}\{Var_{\boldsymbol{u}_j}[f_j(X_j)|\sigma^2_j=1,X_j]\}$, it is easy to see that $C_j$ is a scaling solution adjacent to the \textit{reference variance} proposed by \citet{SR14}, i.e.:
\begin{gather*}
\sigma^2_{\text{ref}}= \text{GM}_{X_j}\left\{Var_{\boldsymbol{u}_j}[f_j(X_j)|\sigma^2_j=1,X_j]\right\}.
\end{gather*}
In fact, the scaling step of Proposition \ref{prop:standardization} can be viewed as a variant of the method by \citet{SR14}, in which the geometric mean is replaced by the expected value. 
However, expectation-based scaling has advantages over the geometric mean approach. For example, the geometric mean approach would return a null scaling constant for linear effects, and more generally for all effects for which $\exists x \in \mathcal{X}$ such that $\pi(x)>0$ and $Var_{\boldsymbol{u}}[f(x)|\sigma^2,X=x]=0$. Additionally, the useful result in Remark \ref{remark:VP} is only achievable thanks to the linearity property of expectation, and therefore does not hold under geometric mean scaling.

\section{IGMRFs}\label{sec:IGMRFs}
In this section, we discuss the steps required for a sensible application of the standardization procedure to effects with IGMRF priors, which are often present in LGMs.

IGMRFs are improper multivariate Gaussian distributions, i.e. their structure matrix has a rank-deficiency $d>0$ and therefore a non-empty null space, denoted by matrix $\boldsymbol{S}$ with $d$ columns, such that $\boldsymbol{QS}=\boldsymbol{0}$. \citet{RH} defined various classes of IGMRFs: in this context, we focus on a subclass of Definition 3.5 of \citet{RH}. 

\begin{definition}[IGMRFs on regular locations on the line]\label{def:IGMRF}
Consider a vector of regular locations denoted by $\boldsymbol{x}=[1,\ldots ,K]^T$. A random vector $\boldsymbol{u}\in \mathbb{R}^K$ is an IGMRF of order $d$ with parameters $\boldsymbol{\mu}$ and $\boldsymbol{Q}$ on regular locations on the line if it has density:
\begin{gather*}
\pi(\boldsymbol{u}|\sigma^2)\propto \exp\left[-\dfrac{1}{2\sigma^2}\left(\boldsymbol{u}-\boldsymbol{\mu}\right)^T\boldsymbol{Q}\left(\boldsymbol{u}-\boldsymbol{\mu}\right)\right] \label{eq:IGMRF1}
\end{gather*}
and $\boldsymbol{QS}_{(d-1)}=\boldsymbol{0}$ where $\boldsymbol{S}_{(d-1)}=[\boldsymbol{x}^0,\boldsymbol{x}^1,\ldots ,\boldsymbol{x}^{d-1}]$ is called a polynomial matrix.
\end{definition}

An effect $f(X)$ from Model \ref{def:model} with an IGMRF prior of order $d$ from Definition \ref{def:IGMRF} can be represented as:
\begin{equation}\label{eq:generic_IGMRF}
\begin{aligned}
f(X)&=\boldsymbol{D}^T(X)\boldsymbol{u};\ \ \ \ \boldsymbol{u}|\sigma^2\sim N(\mathbf{0},\sigma^2\boldsymbol{Q}^*) \text{ where }\boldsymbol{QS}_{(d-1)}=\boldsymbol{0}.
\end{aligned}
\end{equation} 
To understand the role of $\sigma^2$ in such effects, we make use of the decomposition from \citet{RH} (Section 3.4.1):
\begin{gather*}
    \boldsymbol{u}=\boldsymbol{H}_{(d-1)}\boldsymbol{u}+(\boldsymbol{I}-\boldsymbol{H}_{(d-1)})\boldsymbol{u}
\end{gather*}
where $\boldsymbol{u}$ is decomposed in a polynomial trend of degree $d-1$ and a residual part through the \textit{hat matrix} $\boldsymbol{H}_{(d-1)}=\boldsymbol{S}_{(d-1)}[\boldsymbol{S}_{(d-1)}^T\boldsymbol{S}_{(d-1)}]^{-1}\boldsymbol{S}_{(d-1)}^T$. Since $\boldsymbol{Q}\boldsymbol{H}_{(d-1)}=\boldsymbol{0}$, the density of $\boldsymbol{u}$ (Eq. (\ref{eq:IGMRF1}) when $\boldsymbol{\mu}=\boldsymbol{0}$) can be rewritten as:
\begin{gather}\label{eq:u_density}
    \pi(\boldsymbol{u})\propto \exp\left[-\dfrac{1}{2\sigma^2}(\boldsymbol{u}-\boldsymbol{H}_{(d-1)}\boldsymbol{u})^T\boldsymbol{Q} (\boldsymbol{u}-\boldsymbol{H}_{(d-1)}\boldsymbol{u})\right].
\end{gather}

Eq. (\ref{eq:u_density}) clearly shows how $\sigma^2$ does not control the dispersion of the IGMRF process around its mean but rather only the deviation from its polynomial trend of degree $d-1$. Hence, the $\sigma^2$ parameter can only be correctly interpreted as the variance of the process after the removal of the polynomial trend, which can be achieved by imposing the \textit{null space constraints} $\boldsymbol{S}^T_{(d-1)}\boldsymbol{u}=\boldsymbol{0}$. The new process $\boldsymbol{u}|\boldsymbol{S}^T_{(d-1)}\boldsymbol{u}=\boldsymbol{0}$ corresponds to a proper GMRF under constraints, whose variance parameter can be correctly interpreted as a measure of dispersion of the process from its mean, i.e. $\mathbf{0}$. However, the constrained process $\boldsymbol{u}|\boldsymbol{S}^T_{(d-1)}\boldsymbol{u}=\boldsymbol{0}$ is not equivalent to the original (unconstrained) process $\boldsymbol{u}$, since it ``has lost'' its polynomial trend of degree $d-1$. As a consequence, $f(X)$ from Eq. (\ref{eq:generic_IGMRF}) will also be different from $f(X)|\boldsymbol{S}^T_{(d-1)}\boldsymbol{u}=\boldsymbol{0}$. Hence, $\sigma^2$ alone is unable to capture the variance contribution of the original effect, only the one of the constrained version. 

In order to specify processes from Eq. (\ref{eq:generic_IGMRF}) so that variance parameters can measure their variance contribution, a single effect is therefore insufficient and it becomes necessary to use a representation of the process through multiple separate effects, each with its own associated variance parameter. In what follows, we describe how such an alternative representation, denoted by $f_{\text{new}}(X)$, can be conveniently built for two classes of effects frequently employed in LGMs: Section \ref{sec:disc_IGMRFs} considers IGMRF effects for discrete supports (e.g. areal/grid data, time series observed at regular time points); Section \ref{sec:psplines} focuses on P-splines effects, which are used to model smooth effects of continuous covariates.

\subsection{IGMRFs for discrete effects}\label{sec:disc_IGMRFs}
Consider an effect from Eq. (\ref{eq:generic_IGMRF}) for a discrete covariate $X$ with support $[1,\ldots ,K]$ with finite $K$ defined as:
\begin{equation}\label{eq:discrete_IGMRF}
f(X)=\sum_{k=1}^K\mathbb{I}(X=k)u_k; \ \ \ \ \boldsymbol{u}|\sigma^2\sim N(\mathbf{0},\sigma^2\boldsymbol{Q}^*) \text{ where }\boldsymbol{QS}_{(d-1)}=\boldsymbol{0}.
\end{equation}
Eq. (\ref{eq:discrete_IGMRF}) covers various discrete space and time effects, as the ICAR for areal data (\citet{besag1995conditional}) and the RW1/RW2 processes for temporal data. Since $X$ represents space (e.g., grid cell/county) or time (e.g., day/year) in most applications of Eq. (\ref{eq:discrete_IGMRF}), it is reasonable to assume $X\sim\text{Unif}[1,K]$ to reflect equal importance of the values of the support. 

From the previous discussion, it is clear that the specification of the effect from Eq. (\ref{eq:discrete_IGMRF}) is inadequate whenever the interpretation of $\sigma^2$ is desirable for prior specification, e.g. for VP priors. In order to build an appropriate alternative representation, we start by considering the effect under the null space constraints, which we formally denote by $f_r(X)$ where $r$ stands for \emph{residual}:
\begin{equation*}
    f_r(X) = \sum_{k=1}^K\mathbb{I}(X=k)u_k ; \ \ \ \ \boldsymbol{u}|\boldsymbol{S}^T_{(d-1)}\boldsymbol{u}=\boldsymbol{0}, \sigma_r^2\sim N(\mathbf{0},\sigma_r^2\boldsymbol{Q}^*).
\end{equation*}
We know that the null space constraints $\boldsymbol{S}^T_{(d-1)}\boldsymbol{u}=\boldsymbol{0}$ guarantee that the coefficients $\boldsymbol{u}$ have a null polynomial trend of degree $d-1$. In this particular case, this is also true for $f_r(X)$ due to the special choice of basis, i.e.:
\begin{gather}\label{eq:null_trend}
E_X[X^m f_r(X)]=0 \quad m=0,\ldots ,d-1.
\end{gather}
See the proof in Section D of the Supplementary Material.
The result of Eq. (\ref{eq:null_trend}) proves that the imposition of the null space constraints on the original process causes the removal of the polynomial trend. Therefore, $f(X)$ can be nicely reconstructed defining $f_{\text{new}}(X)$ as:
\begin{eqnarray}
f_{\text{new}}(X)=f_t(X)+f_r(X) 
\label{eq:fdiscr_dec}
\end{eqnarray}
where $f_t(X)$ ($t$ stands for \emph{trend}) is designed to model the polynomial effect of degree $d-1$ for covariate $X$, and $\sigma^2_t$ is the associated variance parameter.

In the case of $d=1$, $f_t(X)$ is redundant as the polynomial trend would simply be a constant effect with respect to $X$, whose inclusion would clash with the intercept parameter $\mu$ in the linear predictor, and whose variance contribution will be anyway null. Hence, $f_{\text{new}}(X)=f_r(X)$ and the single $\sigma^2_r$ is sufficient to measure the variance contribution of the effect. For $d=2$ instead, the polynomial term must be a linear effect $f_t(X)=X\beta$ where $\beta|\sigma^2_t\sim N(0,\sigma^2_t)$, so that $f_{\text{new}}(X)=f_t(X)+f_r(X)$. Since IGMRFs of higher order are rarely used in practice, the design of $f_t(X)$ for $d>2$ is discussed in Section E of the Supplementary Material.
In order to obtain the interpretability of $\sigma^2_t$ and $\sigma^2_r$, the two effects must be separately standardized according to Proposition \ref{prop:standardization}. Note how the null space constraints already imply $E_X[f_r(X)]=0$ (i.e. a 0-mean constraint on $f_r(X)$) for every $d>0$: hence, the standardization of $f_r(X)$ is invariant to the fixed/random effect classification. After standardization, it can be proved that (see Section F of the Supplementary Material):
\begin{gather}\label{eq:IGMRF_var}
Var[f_\text{new}(X)|\sigma^2_t,\sigma^2_r]=\sigma^2_t+\sigma^2_r
\end{gather}
where $\sigma^2_t$ controls the polynomial contribution, i.e. the deviation of the polynomial trend from the null mean of the process, while $\sigma^2_r$ controls the residual contribution, i.e. the deviation of the process from its polynomial trend. The process $f_\text{new}(X)$ converges to the original $f(X)$ when $\sigma^2_t\rightarrow\infty$ (i.e., any possible realization of the polynomial trend is equally likely).

\begin{example}[Random walks for temporal effects]\label{ex:3}

The effect of a time covariate with regularly spaced observations is often modelled in LGMs assuming a RW1 or RW2 on the basis coefficients $\boldsymbol{u}$ in Eq. (\ref{eq:discrete_IGMRF}). An RW1 is an IGMRF of order $d=1$ with structure matrix:
\begin{gather}\label{eq:prelim_RW1}
\boldsymbol{Q}_{\text{RW1}}=
\begin{bmatrix}
1 	& -1  &     &     &   \\
-1 	&  2  & -1  &     &   \\
    & \ldots  & \ldots  & \ldots  &   \\
    &     & -1  & 2   & -1\\
    &     &     & -1  &  1\\
\end{bmatrix}.
\end{gather}
Since $d=1$, it is sufficient to impose the null space constraint $\boldsymbol{1}^T\boldsymbol{u}=0$ to obtain a measure of the variance contribution of the effect. After standardization, an RW1 effect can be written as:
\begin{equation}\label{eq:RW1}
f_{\text{new}}(X)=\sum_{k=1}^K\mathbb{I}(X=k)u_k; \ \ \ \ \ \boldsymbol{u}|\boldsymbol{1}^T\boldsymbol{u}=0,\sigma^2\sim N\left(\mathbf{0},\dfrac{\sigma^2_r}{C}\boldsymbol{Q}_{\text{RW1}}^*\right) 
\end{equation}
where $C$ is computed according to Eq. (\ref{eq:scaling_constant}) so that $Var_{X,\boldsymbol{u}}[f_{\text{new}}(X)|\sigma^2_r]=\sigma^2_r$.
Being an IGMRF of order 1, an ICAR effect (\citet{besag1995conditional}) can also be introduced using Eq. (\ref{eq:RW1}), replacing the structure matrix with $\text{diag}(\boldsymbol{W1})-\boldsymbol{W}$ where $\boldsymbol{W}$ must be the adjacency matrix for the given lattice. An RW2 is instead an IGMRF of order $d=2$ with structure matrix:
\begin{gather}\label{eq:prelim_RW2}
\boldsymbol{Q}_{\text{RW2}}=
\begin{bmatrix}
1 	& -2  &  1  &      &    &   &   \\
-2 	&  5  & -4  &  1   &    &   &    \\
1 	&  -4 &  6  &  -4  & 1  &   &    \\
         & \ldots  & \ldots   & \ldots &\ldots & \ldots &   \\
         &      & 1  & -4&  6 & -4& 1 \\
       &      &    & 1 & -4 & 5 & -2\\
      &      &    &  & 1 & -2 & 1\\
\end{bmatrix}.
\end{gather}
In this case, it is necessary to decompose the effect into a linear term and a residual one subject to the null space constraints. After standardization of both terms, the effect can be written as:
\begin{align*}
f_{\text{new}}(X)&=\dfrac{X-E[X]}{\sqrt{Var[X]}}\beta+\sum_{k=1}^K\mathbb{I}(X=k)u_k,\\
\beta|\sigma^2_t&\sim N(0,\sigma^2_t),\\
\boldsymbol{u}|\boldsymbol{S}_{(1)}^T\boldsymbol{u}=\boldsymbol{0},\sigma^2_r&\sim N\left(\mathbf{0},\dfrac{\sigma^2_r}{C}\boldsymbol{Q}_{\text{RW2}}^*\right), 
\end{align*}
so that $Var_{X,\beta,\boldsymbol{u}}[f_{\text{new}}(X)|\sigma^2_t,\sigma^2_r]=\sigma^2_t+\sigma^2_r$. The null space is $\boldsymbol{S}_{(1)}=[\boldsymbol{1}_{K}, (1,\ldots ,K)^T]$.  The scaling constants $C$ for different values of $K$ for both RW1 and RW2 are reported in Section H of the Supplementary Material. 
\end{example}

\subsection{IGMRFs for P-splines}\label{sec:psplines}

Here, we focus on the popular P-spline approach (\citet{EM},  \citet{LB}). P-splines are specified via a B-spline basis defined on a large number of equidistant knots on the support of a continuous covariate $X$, and a random walk on the coefficients, which corresponds to a penalty on the differences between neighboring coefficients. Formally, P-splines can be defined as a special case of Eq. (\ref{eq:generic_IGMRF}):
\begin{equation}\label{eq:cont_IGMRF}
f(X)=\boldsymbol{B}(X)^T\boldsymbol{u}; \ \ \ \ \boldsymbol{u}|\sigma^2\sim N(\mathbf{0},\sigma^2\boldsymbol{Q}^*)  \text{ where }\boldsymbol{QS}_{(d-1)}=\boldsymbol{0}
\end{equation}
where $\boldsymbol{B}(X)=[{B}_1(X),\ldots ,{B}_{K}(X)]^T$ is a basis of $K$ B-spline functions. Since the definition of the basis requires the choice of a finite interval $[m,M]$, it is convenient to assume that $X\sim\text{Unif}(m,M)$.

\begin{figure}[hbt!]
    \centering
\includegraphics[width=\textwidth]{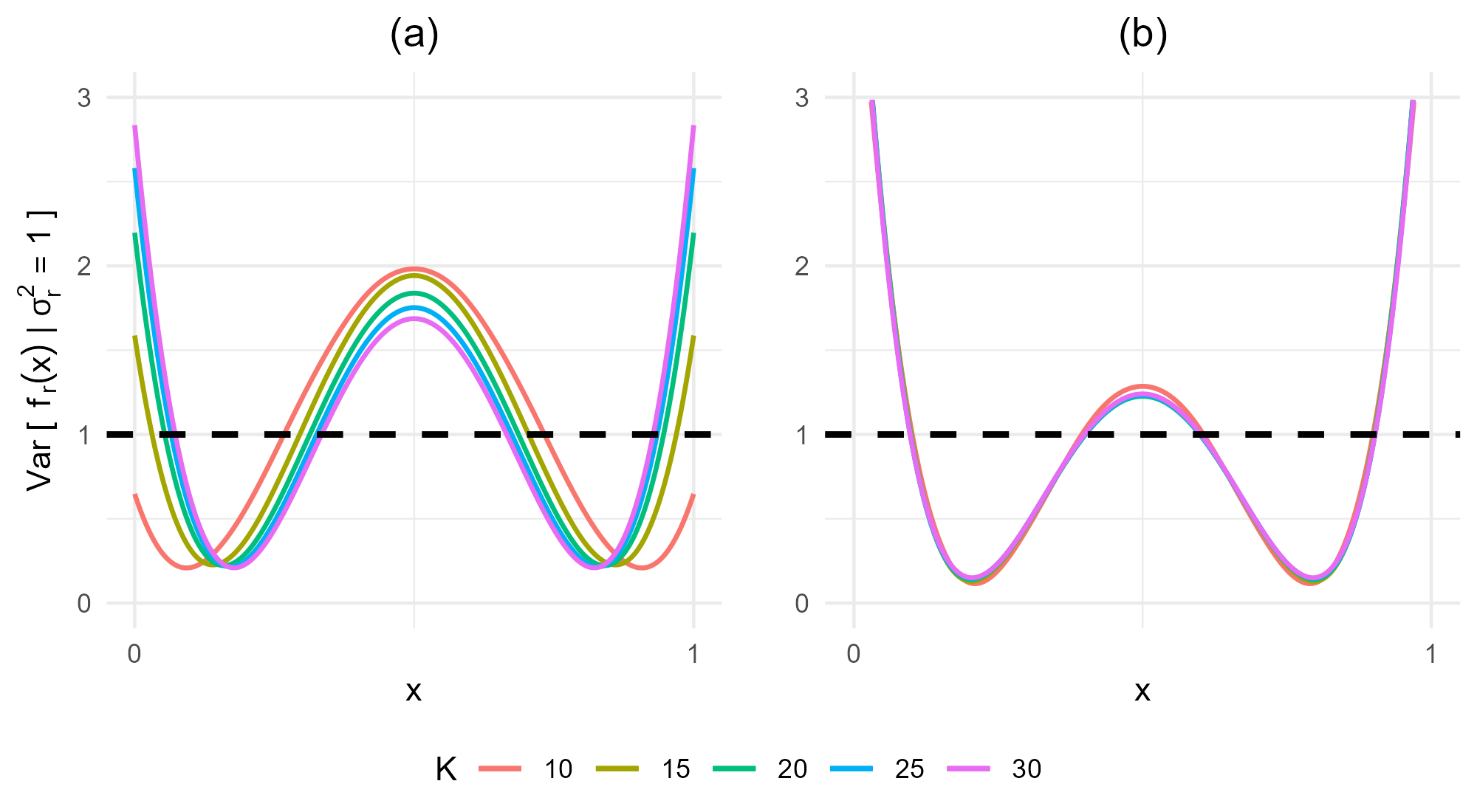}
    \caption{Conditional variance of $f_r(X)$ given $\sigma^2_r=1$ for different values of $K$: (a) before the Q modification, (b) after the Q modification.}
    \label{fig:before_after_Q_mod}
\end{figure}

Defining the concept of variance contribution for a P-spline is challenging as the variance parameter $\sigma^2$ in Eq. (\ref{eq:cont_IGMRF}) does not measure variance contribution directly. A measure of variance contribution is given by the conditional variance, i.e. the variance of $f(X)$ computed under null space constraints at each value of $X$ (see Figure \ref{fig:before_after_Q_mod}). The challenge here is that the latter not only depends on $\sigma^2$ but also on the user-defined number of basis functions $K$  (\citet{ventrucci2016penalized}).
The alternative representation required for the interpretability of the variance parameter could be achieved using Eq. (\ref{eq:fdiscr_dec}) again, with $f_t(X)$ being a polynomial trend effect (see Section \ref{sec:disc_IGMRFs}), and $f_r(X)$ being equal to Eq. (\ref{eq:cont_IGMRF}) subject to $\boldsymbol{S}_{(d-1)}^T\boldsymbol{u}=\boldsymbol{0}$. However, this solution is not viable as the use of a more complex basis is responsible for an important difference with respect to Section \ref{sec:disc_IGMRFs}: the null space constraints do not guarantee a null polynomial trend on $f_r(X)$. As a consequence, the null space constraints are also no longer sufficient to guarantee the 0-mean constraint from Proposition \ref{prop:standardization} on $f_r(X)$: if $f_r(X)$ is to be treated as fixed, it would be necessary to further constrain the process during standardization to guarantee the 0-mean requirement (step 1 of Proposition \ref{prop:standardization}). Additionally, an identifiability issue arises between $f_t(X)$ and $f_r(X)$ for $d>2$, since both terms can potentially capture the polynomial trend in the effect of $X$, as $f_r(X)$ is not appropriately constrained. Consequently, the variance parameters $\sigma^2_t$ and $\sigma^2_r$ can no longer be neatly interpreted as the polynomial and residual contributions to the variance.

In order to solve the issue raised by the more complex basis choice, we propose a procedure whose implementation can guarantee identifiability between the trend and polynomial terms, $f_t(X)$ and $f_r(X)$, and the consequent interpretability of their associated variance parameters. 

The procedure presented below, named \textit{Q modification}, solves the problem by replacing the structure matrix with a new $\widetilde{\boldsymbol{Q}
}$ whose null space $\widetilde{\boldsymbol{S}
}$ is such that the null space constraints $\widetilde{\boldsymbol{S}}^T\boldsymbol{u}=\boldsymbol{0}$ guarantee that $f_r(X)$ has a null polynomial trend of degree $d-1$, i.e. Eq. (\ref{eq:null_trend}). 
The adequate null space $\widetilde{\boldsymbol{S}}$ is:
\begin{gather}\label{eq:general_S_tilde}
\widetilde{\boldsymbol{S}}= \begin{bmatrix}
E_X[X^0 \boldsymbol{B}(X)],\ldots ,E_X[X^{(d-1)} \boldsymbol{B}(X)]
\end{bmatrix}.
\end{gather}
$\widetilde{\boldsymbol{Q}}$ must then be found as the solution of  $\widetilde{\boldsymbol{Q}}\widetilde{\boldsymbol{S}}=\boldsymbol{0}$. We propose a solution that preserves sparsity through the following decomposition:
\begin{gather}\label{eq:decomposition}
    \widetilde{\boldsymbol{Q}} =
(\boldsymbol{\Lambda}\widetilde{\boldsymbol{R}}^*\boldsymbol{\Lambda})^*.
\end{gather}
$\boldsymbol{\Lambda}$ must be a positive diagonal matrix of entries $\boldsymbol{\lambda}=[\lambda_1,\ldots ,\lambda_K]$ and $\widetilde{\boldsymbol{R}}$ a square matrix with the same sparsity and sign structure of $\boldsymbol{Q}$. 
Finding a solution that guarantees $\widetilde{\boldsymbol{Q}}\widetilde{\boldsymbol{S}}=\boldsymbol{0}$ is now equivalent to finding $\widetilde{\boldsymbol{R}}$ such that $\widetilde{\boldsymbol{R}}\boldsymbol{\Lambda}\widetilde{\boldsymbol{S}}=\boldsymbol{0}$. The correct entries of $\widetilde{\boldsymbol{R}}$ are functions of the known elements of $\boldsymbol{Q},\widetilde{\boldsymbol{S}}$, and the unknown $\boldsymbol{\lambda}$. Hence, the new structure matrix $\widetilde{\boldsymbol{Q}}$ is known up to $\boldsymbol{\lambda}$, which can be found minimizing the Kullback-Leibler divergence as:
\begin{gather}\label{eq:KLD}
\widehat{\boldsymbol{\lambda}}=\underset{\boldsymbol{\lambda}>\boldsymbol{0}}{\text{arg min  }}D_{\text{KL}}\;(\;\mathcal{N}_{\widetilde{\boldsymbol{Q}}(\boldsymbol{\lambda})}\;||\;\mathcal{N}_{\boldsymbol{Q}}\;)
\end{gather}
to obtain the closest possible solution to the original $\boldsymbol{Q}$ (\citet{rue2002fitting}).
After the Q modification, the modified P-spline effect can be defined as $f_{\text{new}}=f_{t}(X)+f_{r}(X)$, where $f_{t}(X)$ is defined as detailed in Section \ref{sec:disc_IGMRFs} and:
\begin{gather*}
f_r(X)=\boldsymbol{B}^T(X)\boldsymbol{u};\ \ \ \
\boldsymbol{u}|\widetilde{\boldsymbol{S}}^T\boldsymbol{u}=\boldsymbol{0},\sigma^2_r \sim N(\boldsymbol{0},\sigma^2_r \widetilde{\boldsymbol{Q}}^*) 
\end{gather*}

As in Section \ref{sec:disc_IGMRFs}, $\sigma^2_t$ can be interpreted, after standardization, as the polynomial contribution to the variance, while $\sigma^2_r$ measures the residual contribution. The decomposition in Eq. (\ref{eq:decomposition}) preserves sparsity since $f_r(X)$ can be equivalently defined as $f_r(X)=\boldsymbol{B}^T(X)\boldsymbol{\Lambda}\boldsymbol{u}$ and $\boldsymbol{u}|\widetilde{\boldsymbol{S}}^T\boldsymbol{\Lambda}\boldsymbol{u}=\boldsymbol{0},\sigma^2_r \sim N(\boldsymbol{0},\sigma^2_r \widetilde{\boldsymbol{R}}^*)$,
where $\widetilde{\boldsymbol{R}}$ is by design as sparse as the original structure matrix $\boldsymbol{Q}$.

\begin{example}[Cubic P-spline effects]\label{ex:4}
The most popular choice of P-spline effects consists in the use of a cubic B-spline basis and an RW2 process on the coefficients, so that the structure matrix is $\boldsymbol{Q}_{\text{RW2}}$ from Eq. \ref{eq:prelim_RW2} and $\boldsymbol{Q}_{\text{RW2}}\boldsymbol{S}_{(1)}=\boldsymbol{0}$.
Since it cannot be proven that the null space constraints imply a null polynomial trend on the constrained process $f_r(X)$, the Q modification procedure is implemented to achieve interpretability of the variance parameters. First, the new null space $\widetilde{\boldsymbol{S}}$ must be found applying Eq. (\ref{eq:general_S_tilde}):
\begin{gather*}
    \underset{K\times 2}{\widetilde{\boldsymbol{S}}}=\begin{bmatrix}
        \widetilde{S}_{1,0} & \widetilde{S}_{1,1}\\
        \widetilde{S}_{2,0} & \widetilde{S}_{2,1}\\
        \ldots  & \ldots \\
        \widetilde{S}_{K,0} & \widetilde{S}_{K,1}
    \end{bmatrix}=\begin{bmatrix}
        E_X[\boldsymbol{B}(X)],  E_X[X\boldsymbol{B}(X)]
    \end{bmatrix}.
\end{gather*}
Then, a valid solution for $\widetilde{\boldsymbol{Q}}$ from Eq. (\ref{eq:decomposition}) is found if $\widetilde{\boldsymbol{R}}=\widetilde{\boldsymbol{G}}-\widetilde{\boldsymbol{W}}$  where:
\begin{align}
\widetilde{W}_{k,l}&=
        \begin{cases}
            \dfrac{(l-k) W_{k,l}}{ \lambda_k \lambda_l(\widetilde{S}_{k,0} \widetilde{S}_{l,1}-\widetilde{S}_{k,1} \widetilde{S}_{l,0})}\quad k\neq l\\
            0 \quad\quad\quad \quad\quad \quad\quad \quad\quad \;\; k=l
        \end{cases},\label{eq:new_W_ex_5}\\
      \widetilde{G}_{k,l}&=\mathbb{I}[k=l]
             \dfrac{1}{\lambda_k\widetilde{S}_{k,0}} \left[\sum_{j=1}^K  \widetilde{W}_{k,j}   \lambda_j\widetilde{S}_{j,0}\right]
\label{eq:new_G_ex_5}
\end{align}
and $\boldsymbol{W}=\text{diag}(\text{diag}(\boldsymbol{Q}_{\text{RW2}}))-\boldsymbol{Q}_{\text{RW2}}$.
See the proof in Section G of the Supplementary Material, which also contains the solution for a RW1 on the coefficients. Finally, the matrix $\widetilde{\boldsymbol{Q}}$ that best approximates the original $\boldsymbol{Q}$ is found optimizing $\boldsymbol{\lambda}$ according to Eq. (\ref{eq:KLD}).
After standardization, the P-spline effect can be written as:
\begin{align}
    f_{\text{new}}(X)&=f_t(X)+f_r(X);\\
    f_t(X)&=\frac{X-E[X]}{\sqrt{Var[X]}}\beta\label{eq:adj_PSpline_model_start},\ \ \ \ \beta|\sigma^2_t \sim N(0,\sigma^2_t);\\
    f_r(X)&=\boldsymbol{B}^{T}(X)\boldsymbol{u},
    \ \ \ \
\boldsymbol{u}|\widetilde{\boldsymbol{S}}^T\boldsymbol{u}=\boldsymbol{0},\sigma^2_r\sim N\left(\boldsymbol{0},\dfrac{\sigma^2_r}{C}\widetilde{\boldsymbol{Q}}^*\right).\label{eq:adj_PSpline_model_end}
\end{align}
The scaling constants $C$ for different values of $K$ are reported in Section H of the Supplementary Material.
\end{example}

\begin{figure}[hbt!]
    \centering
\includegraphics[width=\textwidth]{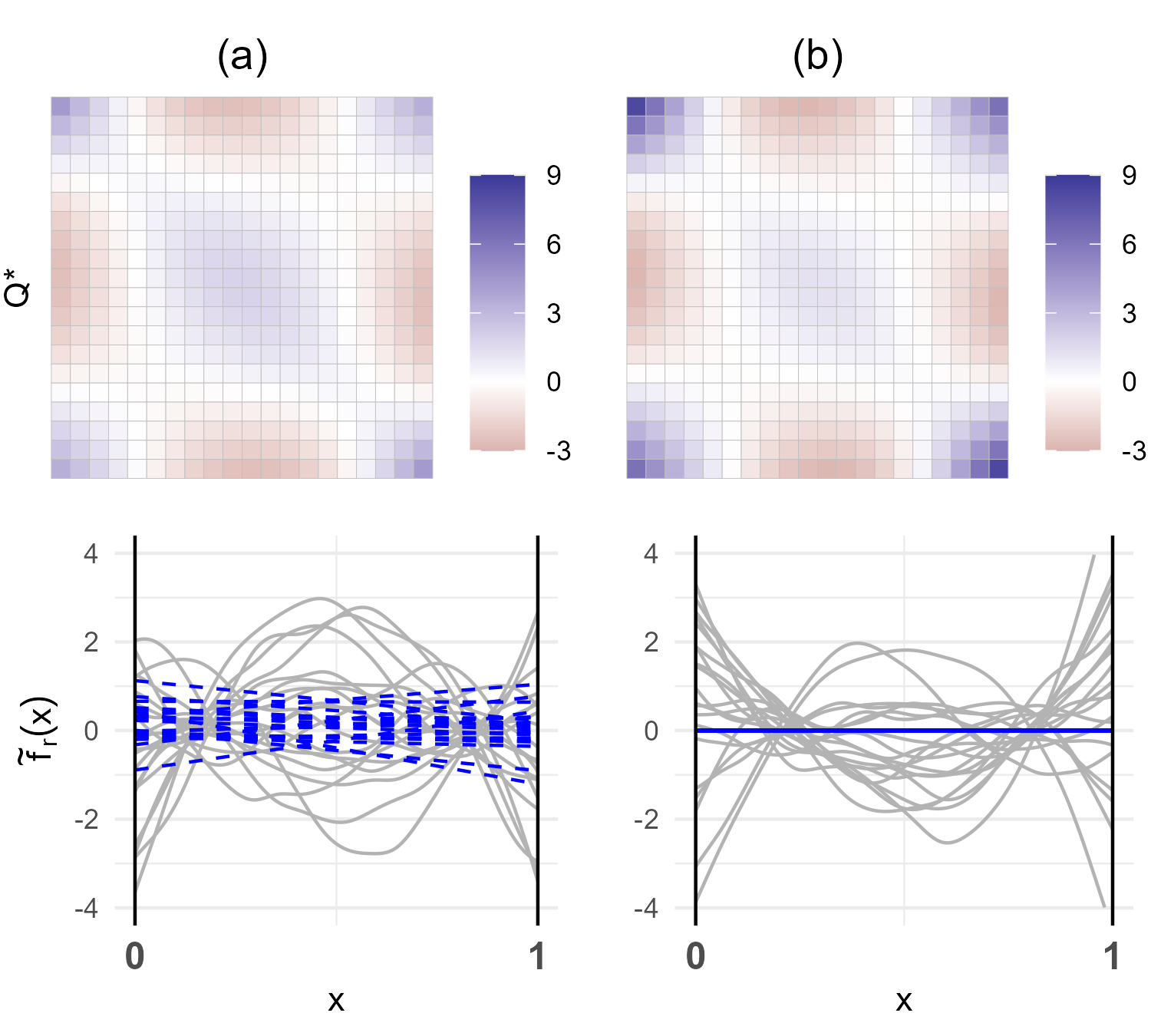}
    \caption{Properties of $f_r(X)$ using (a) $\boldsymbol{Q}_{\text{RW2}}$ and (b) $\widetilde{\boldsymbol{Q}}$ for $K=20$. Top panel: generalized inverse of the structure matrix on the coefficients (each cell represents an element of the matrix). Bottom panel: realizations of $f_r(x)$ (grey) with corresponding linear trends (blue) when $\sigma^2_r=1$.}
    \label{fig:PSpline_RW2_cond_var}
\end{figure}

In order to appreciate the impact of the Q modification, the behaviour of $f_r(X)$ with $K=20$ from Example \ref{ex:4} is compared under the use of $\boldsymbol{Q}_{\text{RW2}}$ and $\widetilde{\boldsymbol{Q}}$ in Figure \ref{fig:PSpline_RW2_cond_var}. First, the top panel shows how the generalized inverse of $\widetilde{\boldsymbol{Q}}$ displays a similar pattern to the one of the original $\boldsymbol{Q}$. The bottom panel of Figure \ref{fig:PSpline_RW2_cond_var} shows realizations of $f_r(x)$ and their linear trends when $\sigma^2_r=1$: the original model has non-null trends despite the null space constraints, while the new model by design removes the linear trends.

Figure \ref{fig:before_after_Q_mod} shows the impact of the Q modification on the shape of the  conditional variance of $f_r(X)$, i.e. $Var_{\boldsymbol{u}}[f_r(X)|\sigma^2=1,X=x]$. It is important to note that the shape of the conditional variance of the process after the Q modification tends to approximate the one of the original model when $K\rightarrow \infty$. Because of this, the conditional variances after the modification (and the standardization) for different values of $K$ are extremely similar (panel b), while this is not the case for the original model (panel a) where we can see that $K$ impacts the conditional variance. 
Therefore, the Q modification has the important advantage of neutralizing the impact of the choice of $K$ on the conditional variance function: in the proposed modified P-spline model, $K$ still controls the local flexibility of the process (i.e. its wiggliness) but not the global shape of the smooth function, which is controlled by the shape of the conditional variance. Finally, note that the impact of the Q modification is less relevant as $K$ grows, since the difference between $\boldsymbol{S}_{(1)}$ and $\widetilde{\boldsymbol{S}}$ decreases.

\section{Empirical results}\label{sec:results}
In this section, we investigate the practical benefits of the standardization procedure and the Q modification through two simulation studies and a survival analysis application on real data.
The models are fitted in R using INLA (\citet{RM}).

\subsection{Simulation study: impact of standardization}\label{sec:scaling_impact}

The aim of the first simulation study is to assess the practical impact of the application of the standardization procedure from Proposition \ref{prop:standardization}.

Standardization allows the variance parameters to match their intuitive interpretation. Hence, it ensures that specifying a prior on a $\sigma^2_j$ implies the same prior on the variance contribution from Definition \ref{def:int_int}. Without standardization instead, the implied prior on the variance contribution could be greatly \textit{distorted}, i.e. different from what was the intended prior. The same is obviously true if a prior specification is chosen on $V$ and $\boldsymbol{\omega}$, which implies the same prior on their interpretations as defined in Remark \ref{remark:VP} only after standardization. The distortion in the prior is a theoretical consequence of incorrect (or lack of) standardization on the model effects. However, the practical impact of standardization can be negligible if: a) the distortion on the implied prior is in itself minimal; b) the information in the data is so overwhelming that the role of the prior becomes irrelevant. With regard to point a), the distortion level depends on the given model specification, as well as on the prior choice on $\boldsymbol{\sigma}$. On the other hand, point b) implies that the importance of standardization grows with the level of prior sensitivity: hence, standardization is particularly important when the data contains limited information about the model. Therefore, we focus here on a simple model for which there is often a weak signal in the data due to non-repeated measurements.

Consider the following simple model:
\begin{gather}\label{eq:sim_1_model}
Y \sim N(\mu+f(X),\sigma^2_\epsilon)
\end{gather}
where $X\sim \text{Unif}[1,25]$ and $f(X)=\sum_{k=1}^{25}\mathbb{I}(X=k)u_k$ is to be considered random (i.e. standardization simplifies to scaling). The coefficients are specified as a first-order random walk (Example \ref{ex:3}), under to the appropriate constraint and three different scaling strategies:
\begin{itemize}
\item Expectation scaling (Proposition \ref{prop:standardization}): $\boldsymbol{u}|\boldsymbol{1}^T\boldsymbol{u}=0,\sigma^2\sim N\left(\boldsymbol{0},\dfrac{\sigma^2}{C}\boldsymbol{Q}_\text{RW1}^*\right)$
\item No scaling: $\boldsymbol{u}|\boldsymbol{1}^T\boldsymbol{u}=0,\sigma^2\sim N\left(\boldsymbol{0},\sigma^2\boldsymbol{Q}_\text{RW1}^*\right)$
\item Geometric mean scaling (\citet{SR14}): $\boldsymbol{u}|\boldsymbol{1}^T\boldsymbol{u}=0,\sigma^2\sim N\left(\boldsymbol{0},\dfrac{\sigma^2}{\sigma^2_{\text{ref}}}\boldsymbol{Q}_\text{RW1}^*\right)$
\end{itemize}
where the scaling constants are equal to $C=4.16$ and $\sigma^2_{\text{ref}}=3.77$.
The parameters of the model in Eq. (\ref{eq:sim_1_model}) are therefore $\mu,\boldsymbol{\sigma}=[\sigma^2,\sigma^2_\epsilon]$. In the Gaussian likelihood case, the VP reparameterization can be applied to $\sigma^2,\sigma^2_\epsilon$ to obtain: 
\begin{gather*}
V=\sigma^2+\sigma^2_\epsilon;\quad \quad \quad \quad \omega=\dfrac{\sigma^2}{\sigma^2+\sigma^2_\epsilon}.
\end{gather*}
We shall also denote by $T$ and $\varphi$ respectively the total variance in $Y$ given the model parameters and the proportion of variance due to $f(X)$: 
\begin{gather*}
T=Var[Y|\mu,\boldsymbol{\sigma}];\quad \quad \quad \quad\varphi =\frac{Var[f(X)|\boldsymbol{\sigma}]}{Var[Y|\mu,\boldsymbol{\sigma}]}.
\end{gather*}
Using VP priors, assumptions about $T$ and $\varphi$ can be introduced through priors on $V$ and $\omega$. This is correct under expectation scaling since $V=T$ and $\omega=\varphi$. Without scaling instead, $T=\sigma^2 C+\sigma^2_\epsilon$ and $\varphi=\frac{\sigma^2 C}{\sigma^2 C+\sigma^2_\epsilon}$: the VP parameters $V$ and $\omega$ are no longer equal to their intuitive interpretations, i.e. $T$ and $\varphi$. The implied priors on $T$ and $\varphi$ would then be distorted in the sense that they would not correctly reflect the user's assumptions about these parameters. For example:
\begin{gather}\label{eq:varphi_induced}
\pi(\varphi )=\pi_\omega\left(\dfrac{\varphi }{\varphi +C-\varphi C}\right) \dfrac{C}{\left[\varphi +C-\varphi C\right]^2}
\end{gather}
since $\omega=\frac{\varphi T}{\varphi T+(1-\varphi)T C}$ (see Section I of the Supplementary Material).
Under geometric mean scaling, the implied priors would also be distorted, since $T=\sigma^2\frac{C}{\sigma^2_{\text{ref}}}+\sigma^2_\epsilon$ and $\varphi=\frac{\sigma^2}{\sigma^2+\sigma^2_\epsilon\sigma^2_{\text{ref}}}$.

To assess the impact of lack or inappropriate standardization, we first compare the distortion on the prior of $T,\varphi$ for three different prior choices, which all implies a symmetric prior on $\omega$:
\begin{enumerate}[label=(\alph*)]
\item \textbf{IG priors}: $\sigma^2,\sigma^2_\epsilon \overset{iid}{\sim} \text{IG}(1,5e-5)$;
\item \textbf{PC priors}: $\sigma^2,\sigma^2_\epsilon \overset{iid}{\sim} \text{PC}_0(3,0.05)$ where $\text{PC}_0$ is a Penalized Complexity prior with null base model, i.e.  $\sigma^2_0=0$ (\citet{S17});
\item \textbf{VP prior}:
$V\sim \text{Jeffreys}, \omega\sim \text{Unif}(0,1)$.
\end{enumerate}

\begin{figure}[hbt!]
    \centering
\includegraphics[width=\textwidth]{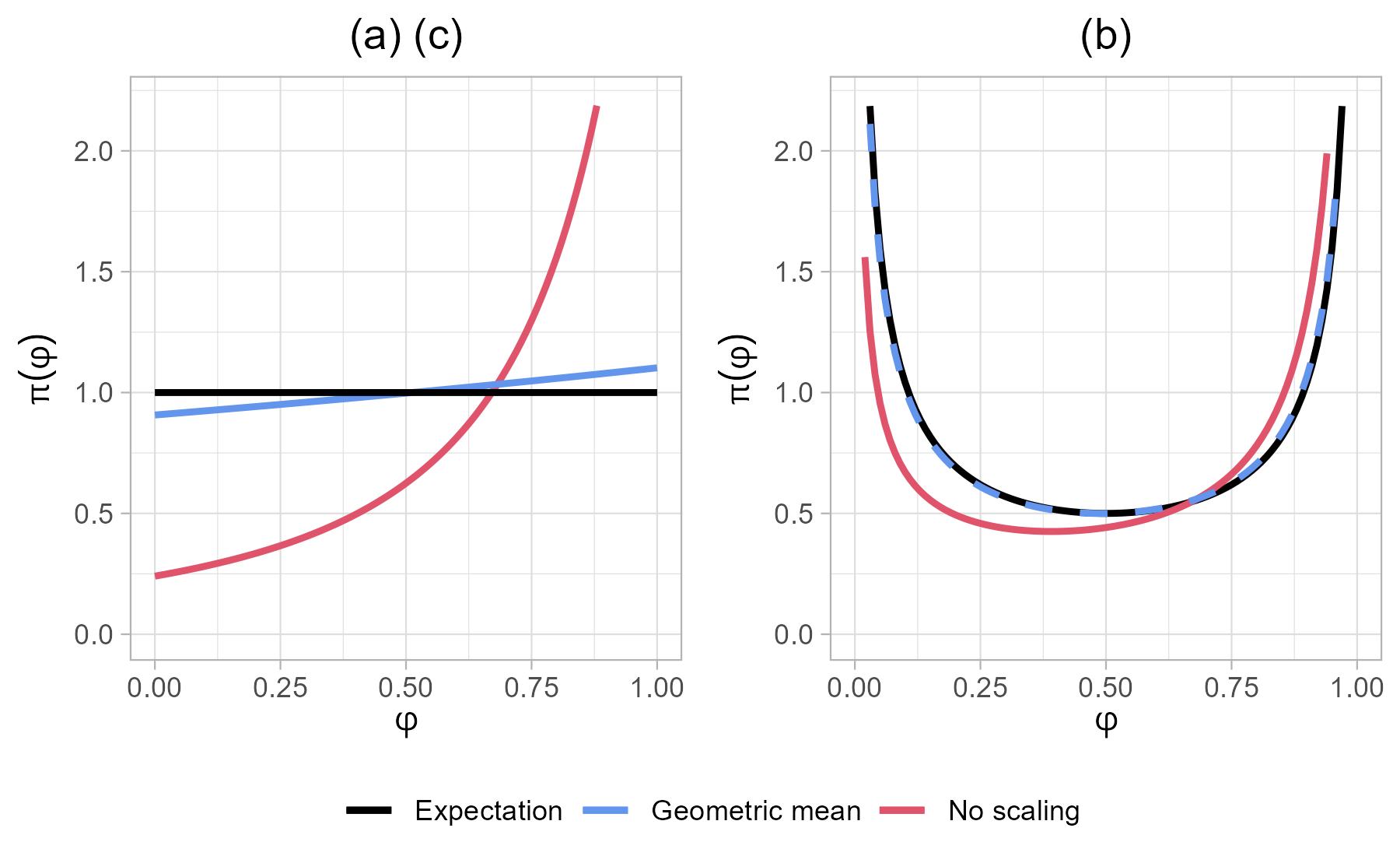}
  \caption{Implied prior on $\varphi $ : (a) IG priors; (b) PC priors; (c) VP prior. The results for the Inverse-Gamma (a) and the VP (c) prior choices are identical and therefore reported in the same panel.}
    \label{fig:RW1_induced_R2_priors}
\end{figure}

Figure
\ref{fig:RW1_induced_R2_priors} reports the density of the implied priors on $\varphi$ for the different scaling strategies. The distortion caused by inappropriate scaling
is assessed comparing the black lines, i.e. the desired priors on $\varphi$, to the other coloured lines, which report the actual implied priors on $\varphi$.
First, it can be noted that the IG and VP prior choices (which return identical results for $\varphi$) seem more sensitive to appropriate scaling than PC priors, since the distortion appears much larger. Secondly, we note that large values of $\varphi $ are favoured under no scaling more than in the desired prior, due to the fact that $C>1$ (the contrary would be true for $C<1$). Thirdly, geometric mean scaling always favours larger values more than the desired prior since by construction $C>\sigma^2_{\text{ref}}$. Nevertheless, the difference between expectation and geometric mean scaling methods appears small, especially for PC priors: this results suggests that geometric mean scaling already removes most of the distortion.

Since we have found non-negligible distortion in the prior, it is necessary to assess whether this distortion affects posterior inference. To do so, a single observation of $Y$ is simulated for each of the $K=25$ locations on $X$, i.e. $x_i=i$ for $i=1,\ldots ,25$. 200 datasets with $N=25$ are generated in this manner using $\mu=0,T=1$ and three different values for $\varphi=$ 0.2, 0.5, 0.8. The datasets are then fitted under the three prior choices and the three scaling strategies. 

\begin{figure}[hbt!]
    \centering
\includegraphics[width=\textwidth]{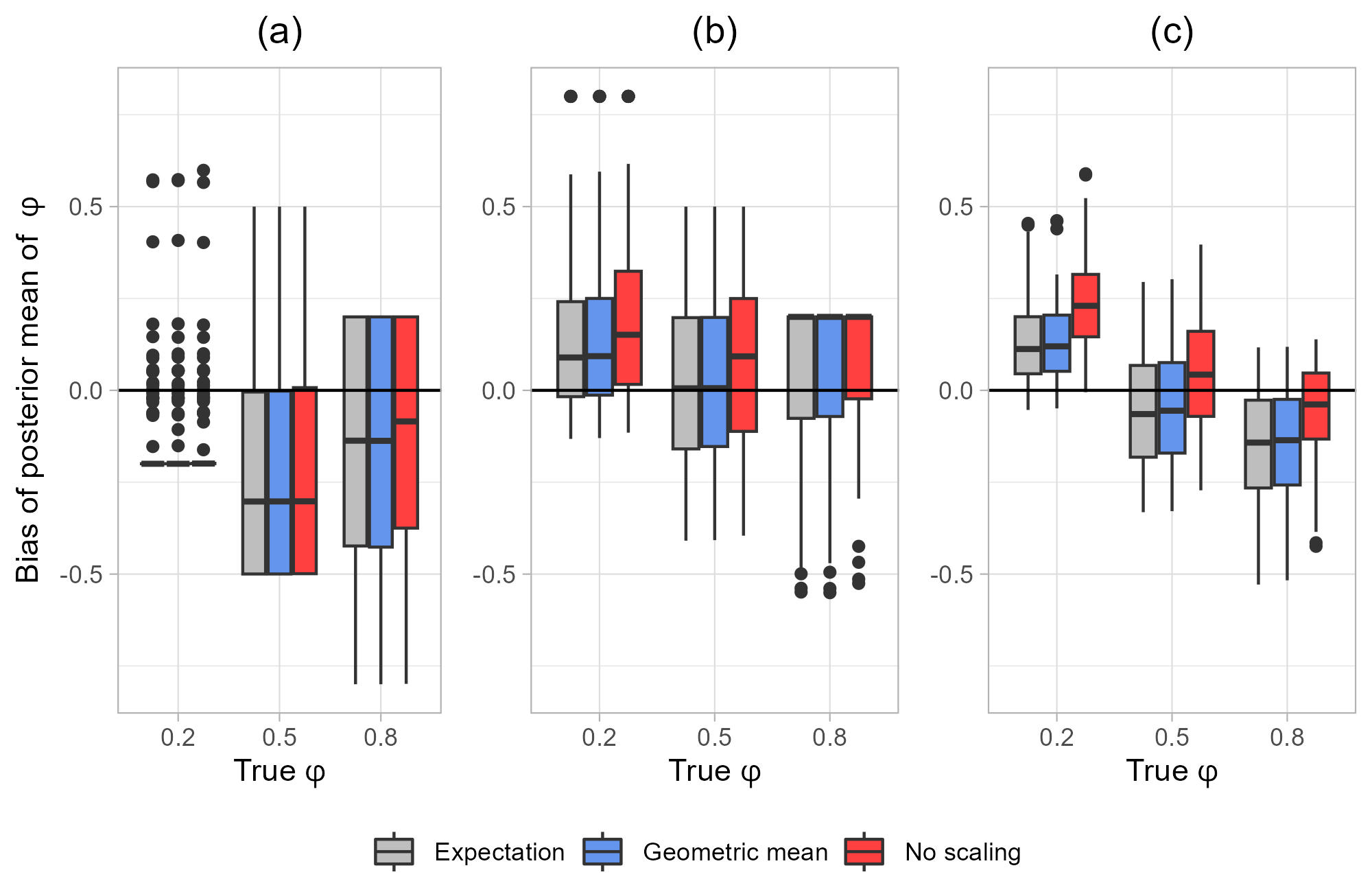}
    \caption{Bias (estimate minus true value) of the posterior mean of $\varphi $ for the local level model under the following prior choices: (a) IG priors; (b) PC priors; (c) VP prior.}
    \label{fig:LLM_logit_R2}
\end{figure}

Figure \ref{fig:LLM_logit_R2} reports the bias of the posterior mean of $\varphi$: there is evidence for a non-negligible difference between the results of a scaled model (either geometric mean or expectation scaling) and an unscaled one. As expected, this difference is larger in the case of the VP prior (c), while the PC priors' choice (b) appears more robust, and even more so the IG prior. In most scenarios, there is no relevant difference between the geometric mean and the expectation scaling strategy. Finally, note how the VP prior performance is arguably the best among the 3 alternatives, with the smallest variance and limited bias.
Similar conclusions can be drawn looking at the bias in the estimates for $T$ in log scale (Figure \ref{fig:LLM_log_T}). Results about coverage rates are reported in Section J of the Supplementary Material.
\begin{figure}[hbt!]
    \centering
\includegraphics[width=\textwidth]{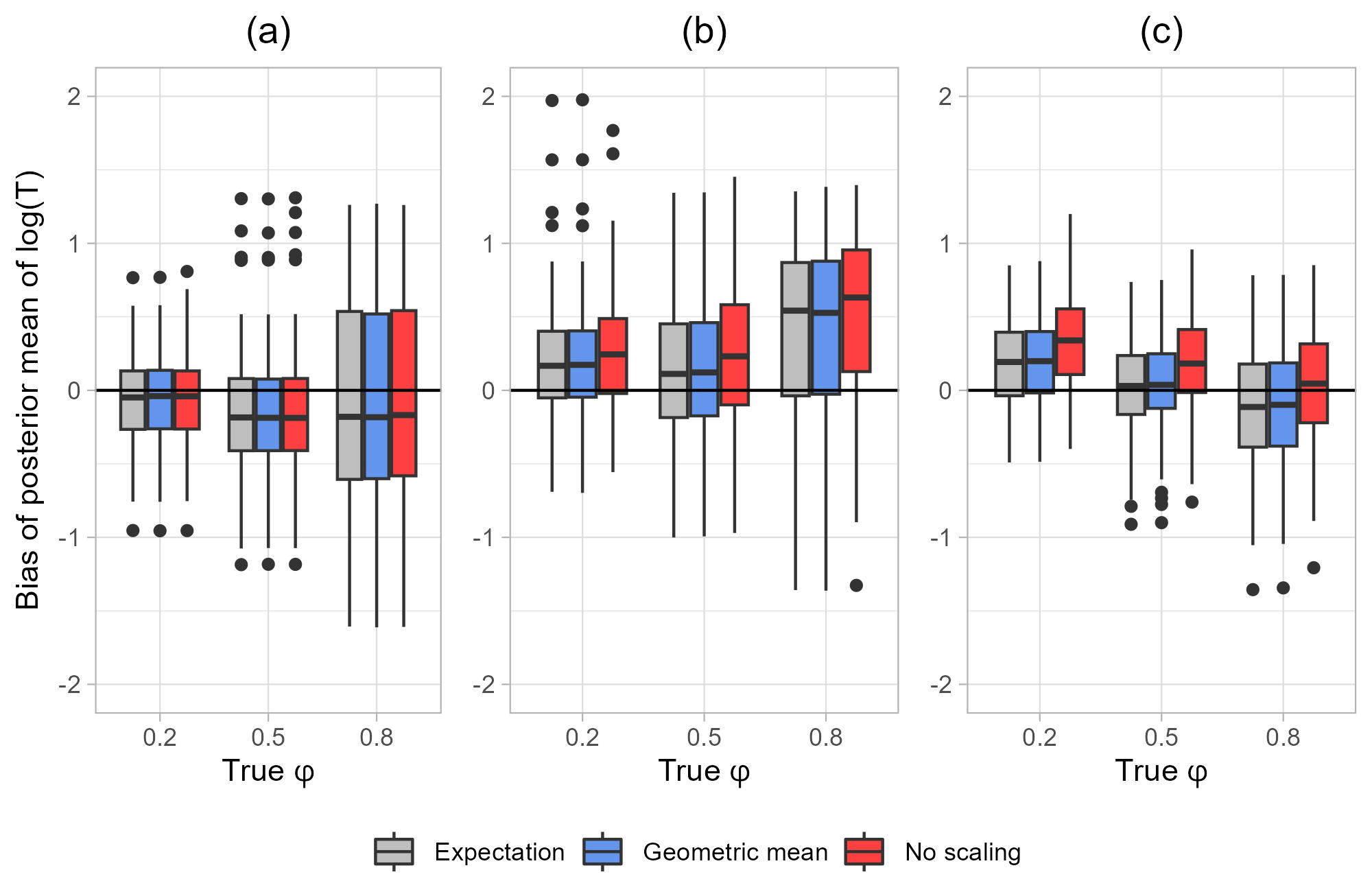}
    \caption{Bias (estimate minus true value) of the posterior mean of $T$ in log scale under the following prior choices: (a) IG priors; (b) PC priors; (c) VP prior.}
    \label{fig:LLM_log_T}
\end{figure}

In summary, the simulation confirmed that VP priors can perform at least as well as the popular PC priors. However, the former seems to suffer more severely from the distortion due to inappropriate standardization than the latter. 
In conclusion, the study just presented highlighted the importance of the scaling step of the standardization procedure; the impact of the 0-mean constraint step has been tested in contexts of Example \ref{ex:2} 
and found to be practically negligible.

\subsection{Simulation study: impact of Q modification}\label{sec:Qmod_impact}
The goal of this second simulation study is to investigate the identifiability issue found in the context of P-spline and assess the practical relevance of applying the proposed Q modification (Section \ref{sec:psplines}).

Consider a Gaussian likelihood model with a smooth effect for continuous covariate $X\sim \text{Unif}(0,1)$ where $Y\sim  N(f_t(X)+f_r(X),\sigma^2_\epsilon)$, $f_t(X)=(X-0.5)\sqrt{12}\beta$ is the linear part and $f_r(X)=\text{cos}(2\pi X)$ is the non-linear one. Let $\varphi$ denote the proportional contribution of the non-linear effect to the variance in the linear predictor:$
\varphi=\frac{Var_X[f_r(X)]}{Var_X[f_t(X)+f_r(X)]}$.
200 datasets are generated with $N=300$ observations and $\sigma^2_\epsilon=1,\beta=\sqrt{0.5}$ so that the variance contribution of $f_t(X)$ and $f_r(X)$ is equal to 0.5, and so is $\varphi$.
\begin{figure}[hbt!]
    \centering
\includegraphics[width=\textwidth]{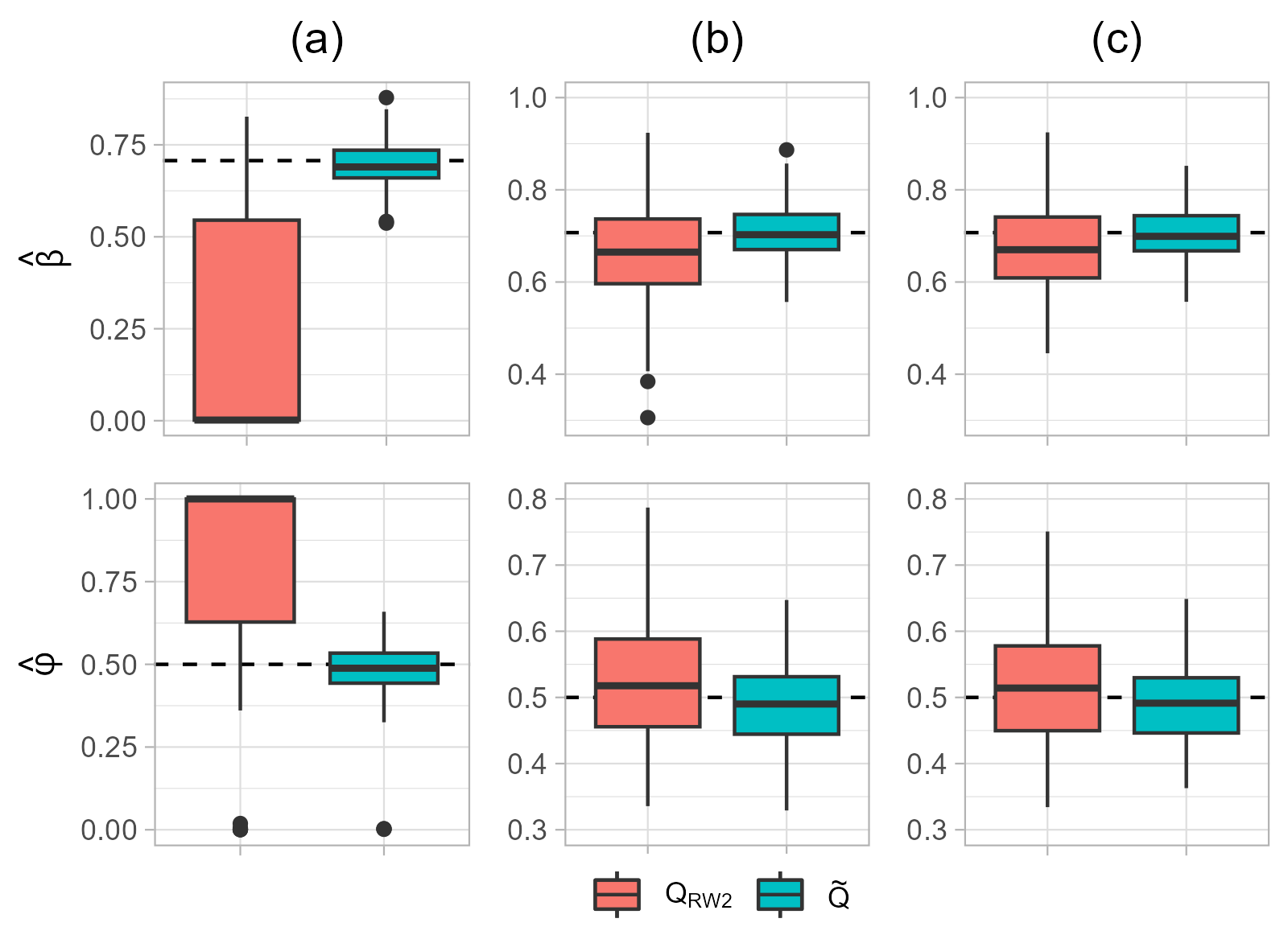}
   \caption{Posterior means of $\beta$ and $\varphi$ (along with their true values in dashed lines) under the choice of the original $\boldsymbol{Q}_{\text{RW2}}$ or its modified version $\widetilde{\boldsymbol{Q}}$: (a) IG priors; (b) PC priors; (c) VP prior.}
    \label{fig:Q_mod_results}
\end{figure}

The response is then fitted using a Gaussian likelihood model with the modified version of the P-spline effect from Equations \ref{eq:adj_PSpline_model_start}-\ref{eq:adj_PSpline_model_end} with $K=10$ (hence, $C=0.835$). The practical impact of the Q modification is evaluated comparing the results to a model using the original prior on the coefficients, that is, $\boldsymbol{u}|\boldsymbol{S}_{(1)}^T\boldsymbol{u}=\boldsymbol{0},\sigma^2_r\sim N\left(\boldsymbol{0},\frac{\sigma^2_r}{C}\boldsymbol{Q}^*_{\text{RW2}}\right)$ for $C=1.432$. Both models are fitted with three different priors:
\begin{enumerate}[label=(\alph*)]
\item \textbf{IG priors}: $\sigma^2_t,\sigma^2_r,\sigma^2_\epsilon \overset{iid}{\sim} \text{IG}(1,5e-5)$;
\item \textbf{PC priors}: $\sigma^2_t,\sigma^2_r,\sigma^2_\epsilon \overset{iid}{\sim} \text{PC}_0(3,0.05)$;
\item \textbf{VP prior}:
$V=\sigma^2_t+\sigma^2_r \sim \text{Jeffreys}, \omega=\sigma^2_t/V\sim \text{Unif}(0,1),\sigma^2_\epsilon \sim \text{IG}(1,5e-5)$.
\end{enumerate}
To evaluate the impact on the estimation of the linear/non-linear contributions, Figure \ref{fig:Q_mod_results} reports the posterior means of $\beta$ and $\varphi$. The results for prior choice (a) highlight the identifiability issue in the original the P-spline effect: without the Q modification, the estimates can be highly biased since the linear contribution can be partially or totally absorbed by the non-linear term. For the prior choices (b) and (c), the gap between the two models reduces, although the Q modification still improves the estimates for $\beta$ and $\varphi$, both in terms of bias and variance.

The Q modification does not seem to affect fitting performance, as summaries of residuals and conditional predictive ordinates remain unchanged (see Section K of the Supplementary Material).
Further simulations with different linear/non-linear contribution ratios show similar results. As $K$ grows, the Q modification becomes less important and the difference in estimation performance becomes negligible for $K>25$.

\subsection{Case study: leukaemia in North West England}\label{sec:application}
We consider the dataset analysed by \citet{henderson2002modeling}, already studied in \citet{kneib2007mixed} and \citet{SR14}. The dataset contains survival times of $N=1043$ patients, diagnosed with adult acute myeloid leukaemia between 1982 and 1998 in the North West England (UK). The following covariates are reported for each patient: $Age$, $Wbc$ (white blood cells count at diagnosis), $Tpi$ (Townsend social deprivation index), $Sex$, $District$ (district of residence).\citet{martino2011approximate} illustrated how a survival analysis can be carried out in INLA (i.e. via an LGM) under the assumption of a piecewise log-constant proportional hazard model (\citet{breslow1972disussion}).
In this case, the linear predictor of the model (i.e. the log-hazard function) is specified as:
\begin{gather*}
\eta=\mu+f_1(Age)+f_2(Wbc)+f_3(Tpi)+f_4(Sex)+f_T(Time)+f_S(District)
\end{gather*}
where $Time$ is a discretization of the survival time in $K_T=27$ intervals.

The effects $f_1(Age)$, $f_2(Wbc)$, $f_3(Tpi)
$ are modelled as P-spline effects with $K=50$ basis functions (Equations \ref{eq:adj_PSpline_model_start}-\ref{eq:adj_PSpline_model_end}). We denote by $\sigma^2_{t1},\sigma^2_{t2},\sigma^2_{t3}$ the variance parameters of the trend terms, and by $\sigma^2_{r1},\sigma^2_{r2},\sigma^2_{r3}$ the parameters of the residual terms. $f_4(Sex)$ is set to a group effect (Example \ref{ex:2}), while a Besag model (\citet{besag1995conditional}) based on the adjacency matrix of the districts is used for $f_S(District)$, and a first-order random walk is chosen for $f_T(T)$ (Eq. (\ref{eq:RW1})): these effects are respectively associated with variance parameters $\sigma^2_{4},\sigma^2_{S},\sigma^2_{T}$. All effects are treated as fixed, a discrete Uniform distribution is assumed for $Sex,Time,District$, and a continuous one for $Age, Wbc,Tpi$, on their respective empirical ranges. On the basis of this distributional choice, the necessary 0-mean and null space constraints are imposed on the effects.

In terms of prior, the VP reparameterization from Definition \ref{def:VP} is applied to all the 9 variance parameters and a simple HD prior is assumed for convenience: $V\sim \text{Jeffreys}$ and $ \boldsymbol{\omega}\sim \text{Dir}(1,\ldots ,1)$. A more thoughtful prior design could entail, for instance, the use of $\text{PC}_0$ priors on the proportions $\sigma^2_{rp}(\sigma^2_{tp}+\sigma^2_{rp})^{-1}$ for $p=1,2,3$ to penalize non-linearity.

\begin{figure}
    \centering
\includegraphics[width=\textwidth]{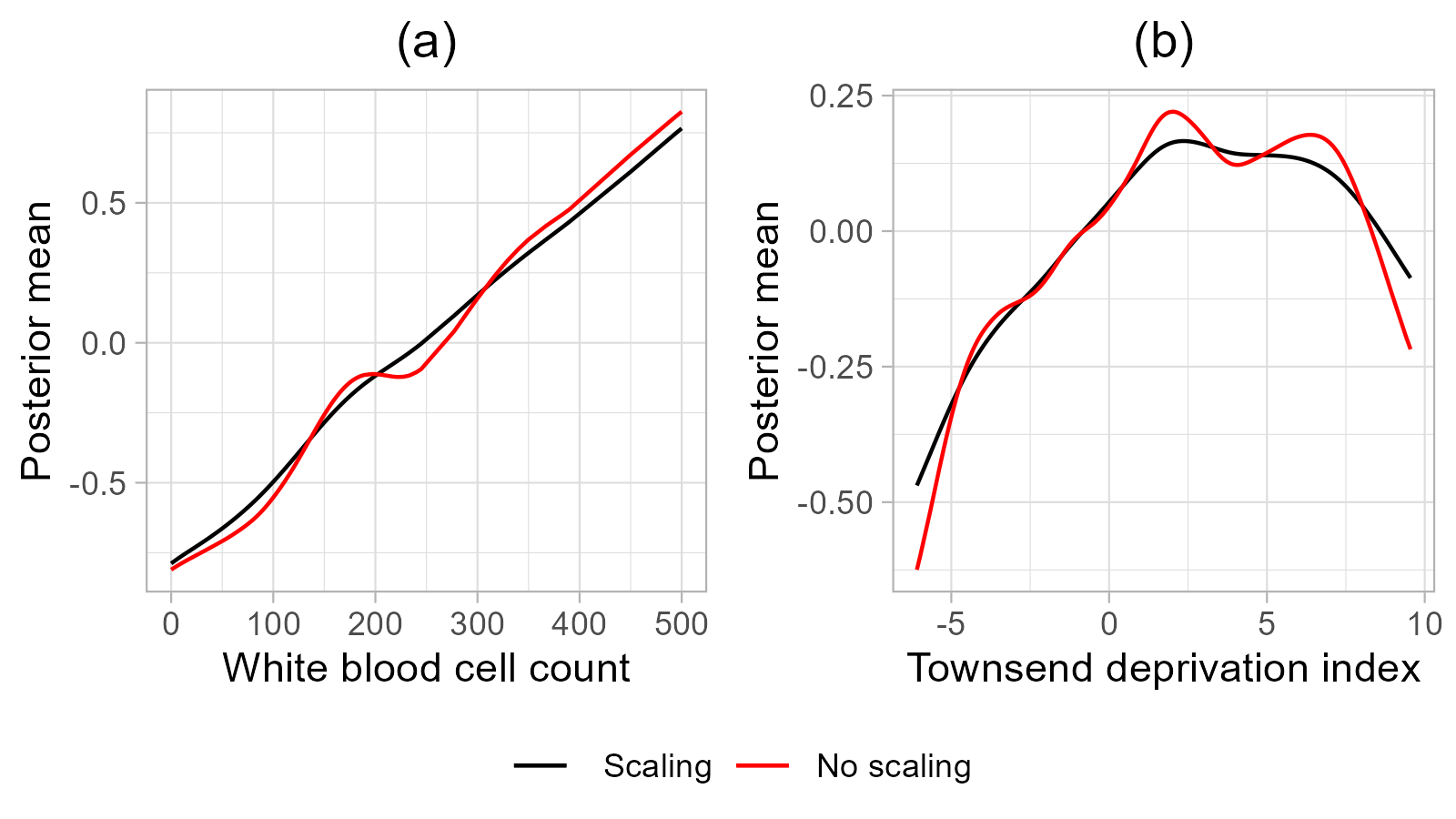}
    \caption{Posterior means of (a) $f_2(Wbc)$ and (b) $f_3(Tpi)$ before (red) and after expectation scaling (black).}
    \label{fig:case_study}
\end{figure}

Figure \ref{fig:case_study} reports the posterior mean of the effects for $Wbc,Tpi$ before and after the application of the scaling step of the standardization procedure: lack of appropriate scaling significantly affects the smoothness of the estimated functions, making them more wiggly than necessary. The remaining effects are not significantly affected by scaling.
Overall, the scaled solution reports results that are coherent with past analyses of the dataset, suggesting for example a linear trend for both $Age$ and $Wbc$ (\citet{kneib2007mixed}).
Very similar results are obtained if only the linear effects are standardized
(i.e. standardizing the covariates), which can be considered the default approach usually adopted in practice. The geometric mean scaling could not be applied to all the effects in the model due to the presence of the linear ones; however, the results are also very similar when expectation scaling is applied to linear effects and the geometric mean one is used on the remaining effects. Finally, the impact of scaling is greatly reduced if independent $\text{PC}$ priors are used instead of the VP prior, proving once more the robustness of the PC framework.

Additional results and R code scripts to replicate the analysis are available at \url{https://github.com/LFerrariIt/scaleGMRF/tree/master/Leuk_application}.

\section{Discussion}
\label{sec:discussion}
In this paper, we aimed at extending the applicability of VP priors, which requires an alignment between the variance parameters of the effects and their corresponding variance contribution. To formally define the ``variance contribution'' of an effect, we have recognized how this concept is differently interpreted by the user according to what is the parameter of the effect that is of inferential interest (i.e. fixed/random effects classification) and therefore decided to use the \emph{expected variance of interest} as the intuitive interpretation of a variance parameter. We have then identified which conditions are necessary to guarantee the alignment, and used them to define a standardization procedure to be applied to all effects according to their fixed/random nature. Simulations confirm the importance of standardization, especially scaling, for the correct use of VP priors. Our scaling method, while similar to \citet{SR14}, is arguably preferable due to the result in Remark \ref{remark:VP}.

This work also considered how the concept of variance contribution can be measured for IGMRF effects, which can be achieved through a representation that separates the polynomial and residual components into distinct effects with individual variance parameters. While this separation is straightforward in simpler cases, P-spline enecessitates an additional ``Q modification'' to the IGMRF structure matrix for the removal of identifiability issues. The Q modification utility goes beyond P-splines and can also be used to adjust other complex IGMRFs, such as those created using Kronecker products of univariate IGMRFs to model interactions.

We hope this work will lead to further development of the HD framework by \citet{F20}, now that a standardization procedure to intuitively interpret the variance parameters of both fixed and random effects is available.  
Additionally, the possibility to apply VP priors to models including smooth effects on covariates via P-splines may lead to the extension of the R2D2 approach to the context of non-linear sparse regression (\citet{wei2020sparse}).
In conclusion, we believe that this contribution may facilitate broader exploitation of the VP prior approach, which allows an easier and more accurate reflection of prior beliefs.

\paragraph{Funding}
The authors were supported by the European Union under the NextGeneration EU Programme within the Plan “PNRR - Missione 4 “Istruzione e Ricerca” - Componente C2 Investimento 1.1 “Fondo per il Programma Nazionale di Ricerca e Progetti di Rilevante Interesse Nazionale (PRIN)” by the Italian Ministry of University and Research (MUR), Project title: “METAbarcoding for METAcommunities: towards a genetic approach to community ecology (META2) ”, Project code: 2022PA3BS2 (CUP E53D23007580006), MUR D.D. financing decree n. 1015 of 07/07/2023

\newpage
\bibliography{biblio}

\newpage

\appendix

\begin{center}
\LARGE{\textbf{Supplementary Material}}
\end{center}

\setcounter{section}{0}
\renewcommand\thesection{\Alph{section}}
\section{Proof of Proposition \ref{prop:standardization}}\label{sm:proposition}
Proposition \ref{prop:standardization} can be proved in multiple steps. Recall that the $j^{th}$ effect from Model 1 is defined as $f_j(X_j)=\mathbf{D}_j(X_j)^T\mathbf{u}_j$. First, it can be noticed that the definition of intuitive interpretation requirement for the fixed effects (Eq. (\ref{eq:fe_int_req})) can be rewritten as a difference between two terms if the variance is written in terms of difference of expectations:
\begin{align*}
E_{\boldsymbol{u}_j}\{Var_{X_j}[f_j(X_j)|\boldsymbol{u}_j]|\sigma^2_j\}
&=E_{\boldsymbol{u}_j}\{E_{X_j}[f^2_j(X_j)|\boldsymbol{u}_j]-E_{X_j}^2[f_j(X_j)|\boldsymbol{u}_j]|\sigma^2_j\}\\
&=E_{\boldsymbol{u}_j}\{E_{X_j}[f^2_j(X_j)|\boldsymbol{u}_j]|\sigma^2_j\}-E_{\boldsymbol{u}_j}\{E_{X_j}^2[f_j(X_j)|\boldsymbol{u}_j]|\sigma^2_j\}
\end{align*}
At this stage, the order of integration in the first term on the right-hand side can be changed as long as $E_{\boldsymbol{u}_j}\{E_{X_j}[f^2_j(X_j)|\boldsymbol{u}_j]|\sigma^2_j\}$ is finite (Fubini-Tonelli theorem). Inverting the expectations, the first term becomes equal to the marginal variance given $\sigma^2_j$. The second term on the right-hand side can also be rewritten as a variance, noting that $E_{\boldsymbol{u}_j}\{E_{X_j}[f_j(X_j)|\boldsymbol{u}_j]|\sigma^2_j\}=0$:
\begin{align*}
E_{\boldsymbol{u}_j}[Var_{X_j}[f_j(X_j)|\boldsymbol{u}_j]|\sigma^2_j]
&=E_{X_j}\{E_{\boldsymbol{u}_j}[f^2_j(X_j)|X_j]|\sigma^2_j\}-E_{\boldsymbol{u}_j}\{E_{X_j}^2[f_j(X_j)|\boldsymbol{u}_j]|\sigma^2_j\}\\
&=Var_{X_j,\boldsymbol{u}_j}[f_j(X_j)|\sigma^2_j]-Var_{\boldsymbol{u}_j}\{E_{X_j}[f_j(X_j)|\boldsymbol{u}_j]|\sigma^2_j\}
\end{align*}

Hence, a 0-mean constraint on $j=1,...,L$ (step 1 of Proposition \ref{prop:standardization}) guarantees that $Var_{\boldsymbol{u}_j}\{E_{X_j}[f_j(X_j)|\boldsymbol{u}_j]|\sigma^2_j\}=0,\;j=1,...,L$ so that:
\begin{gather*}
E_{\boldsymbol{u}_j}[Var_{X_j}[f_j(X_j)|\boldsymbol{u}_j]|\sigma^2_j]
=Var_{X_j,\boldsymbol{u}_j}[f_j(X_j)|\sigma^2_j] \quad j=1,...,L
\end{gather*}
Finally, if the original effect is replaced with $\widetilde{f}_j(X_j)=f_j(X_j)/\sqrt{C}_j$ (step 2 of Proposition \ref{prop:standardization}), then we obtain that:
\begin{eqnarray*}
Var_{X_j,\boldsymbol{u}_j}\left[\widetilde{f}_j(X_j)|\sigma^2_j\right]&=&\dfrac{1}{C_j}Var_{X_j,\boldsymbol{u}_j}\left[f_j(X_j)|\sigma^2_j\right]\\
&=&\dfrac{1}{C_j}E_{X_j}\left\{Var_{\boldsymbol{u}_j}\left[f_j(X_j)|\sigma^2_j\right]\right\}\\
&=&\dfrac{\sigma^2_j}{C_j} E_{X_j}\left[\boldsymbol{D}^T_j(X_j)\boldsymbol{Q}^*_j\boldsymbol{D}_j(X_j)\right]\\
&=&\dfrac{\sigma^2_j C_j}{C_j}=\sigma^2_j.
\end{eqnarray*}

Hence, we have found that Proposition \ref{prop:standardization} ensures that:
\begin{gather*}
\sigma^2_j=Var_{X_j,\boldsymbol{u}_j}[\widetilde{f}_j(X_j)|\sigma^2_j]
\quad j=1,...,J
\end{gather*}
but also that:
\begin{gather*}
\sigma^2_j=E_{\boldsymbol{u}_j}\{Var_{X_j}[\widetilde{f}_j(X_j)|\boldsymbol{u}_j]|\sigma^2_j\}
\quad j=1,...,L
\end{gather*}
which proves that the two steps of Proposition \ref{prop:standardization} are sufficient to guarantee Definition \ref{def:int_int}.
\section{Proof of Remark \ref{remark:VP}}\label{sm:remark}
Remark \ref{remark:VP} can be proven showing that $E_{\boldsymbol{\theta}}\{Var_{\boldsymbol{X},\boldsymbol{u}_1,...,\boldsymbol{u}_J}[\eta|\mu,\boldsymbol{\theta}]|\boldsymbol{\sigma}\}=\sum_{j=1}^J\sigma^2_j$.

First, we can note that:
\begin{align*}
E_{\boldsymbol{\theta}}\{Var_{\boldsymbol{X},\boldsymbol{u}_1,...,\boldsymbol{u}_J}[\eta|\mu,\boldsymbol{\theta}]|\boldsymbol{\sigma}\}&=E_{\boldsymbol{\theta}}\{Var_{\boldsymbol{X},\boldsymbol{u}_1,...,\boldsymbol{u}_J}[\sum_{j=1}^J f_j(X_j)|\boldsymbol{\theta}]|\boldsymbol{\sigma}\}
\end{align*}
Secondly, we can rewrite this expression as: 
\begin{align*}
E_{\boldsymbol{u}_1,...,\boldsymbol{u}_L}\left\{Var_{\boldsymbol{X},\boldsymbol{u}_{L+1},...,\boldsymbol{u}_J}\left[\sum_{j=1}^J f_j(X_j)|\boldsymbol{u}_{1},...,\boldsymbol{u}_L,\sigma^2_{L+1},...,\sigma^2_J\right]|\sigma^2_1,...,\sigma^2_L\right\}
\end{align*}
This expression can be rewritten more concisely using the notation $\boldsymbol{U}_F=[\boldsymbol{u}_1,...,\boldsymbol{u}_L]$, $\boldsymbol{U}_R=[\boldsymbol{u}_{L+1},...,\boldsymbol{u}_J]$,$\boldsymbol{\sigma}_F=[\sigma^2_1,...,\sigma^2_L]$, $\boldsymbol{\sigma}_R=[\sigma^2_{L+1},...,\sigma^2_J]$ as:
\begin{gather*}
E_{\boldsymbol{U}_F}\left\{Var_{\boldsymbol{X},\boldsymbol{U}_R}\left[\sum_{j=1}^J f_j(X_j)|\boldsymbol{U}_F,\boldsymbol{\sigma}_R\right]|\boldsymbol{\sigma}_F\right\}
\end{gather*}
Then, we can write the argument of the expectation using the law of total variance as:
\begin{align*}
 Var_{\boldsymbol{X},\boldsymbol{U}_R}\left[\sum_{j=1}^J f_j(X_j)|\boldsymbol{U}_F,\boldsymbol{\sigma}_R\right]&=E_{\boldsymbol{X}}\left\{Var_{\boldsymbol{U}_R}\left[\sum_{j=1}^J f_j(X_j)|\boldsymbol{X},\boldsymbol{U}_F,\boldsymbol{\sigma}_R\right]|\boldsymbol{U}_F,\boldsymbol{\sigma}_R\right\}\\
 &+Var_{\boldsymbol{X}}\left\{E_{\boldsymbol{U}_R}\left[\sum_{j=1}^J f_j(X_j)|\boldsymbol{X},\boldsymbol{U}_F,\boldsymbol{\sigma}_R\right]|\boldsymbol{U}_F,\boldsymbol{\sigma}_R\right\}\\
  &=E_{\boldsymbol{X}}\left\{Var_{\boldsymbol{U}_R}\left[\sum_{j=L+1}^J f_j(X_j)|\boldsymbol{X},\boldsymbol{\sigma}_R\right]|\boldsymbol{\sigma}_R\right\}\\
 &+Var_{\boldsymbol{X}}\left[\sum_{j=1}^L f_j(X_j)|\boldsymbol{U}_F\right]\\
  &=\sum_{j=L+1}^J E_{\boldsymbol{X}}\left\{Var_{\boldsymbol{U}_R}\left[f_j(X_j)|\boldsymbol{X},\boldsymbol{\sigma}_R\right]|\boldsymbol{\sigma}_R\right\}\\
 &+Var_{\boldsymbol{X}}\left[\sum_{j=1}^L f_j(X_j)|\boldsymbol{U}_F\right]\\
  &=\sum_{j=L+1}^J E_{\boldsymbol{X}}\left\{E_{\boldsymbol{U}_R}\left[f_j^2(X_j)|\boldsymbol{X},\boldsymbol{\sigma}_R\right]\right\}\\
 &+Var_{\boldsymbol{X}}\left[\sum_{j=1}^L f_j(X_j)|\boldsymbol{U}_F\right]
\end{align*}
If a 0-mean constraint is imposed on the $j=1,...,L$ effects, then:
\begin{gather*} 
Var_{\boldsymbol{X}}\left[\sum_{j=1}^L f_j(X_j)|\boldsymbol{U}_F\right]=E_{\boldsymbol{X}}\left\{\left[\sum_{j=1}^L f_j(X_j)\right]^2|\boldsymbol{U}_F\right\}
\end{gather*}
so that:
\begin{align*}
 Var_{\boldsymbol{X},\boldsymbol{U}_R}\left[\sum_{j=1}^J f_j(X_j)|\boldsymbol{U}_F,\boldsymbol{\sigma}_R\right]
   &=\sum_{j=L+1}^J E_{\boldsymbol{X}}\left\{E_{\boldsymbol{U}_R}\left[f_j^2(X_j)|\boldsymbol{X},\boldsymbol{\sigma}_R\right]\right\}\\
 &+E_{\boldsymbol{X}}\left\{\left[\sum_{j=1}^L f_j(X_j)\right]^2|\boldsymbol{U}_F\right\}
\end{align*}
If we consider again $E_{\boldsymbol{U}_F}\left\{Var_{\boldsymbol{X},\boldsymbol{U}_R}\left[\sum_{j=1}^J f_j(X_j)|\boldsymbol{U}_F,\boldsymbol{\sigma}_R\right]|\boldsymbol{\sigma}_F\right\}$, it can be written as:
\begin{align*}
E_{\boldsymbol{U}_F}\left\{Var_{\boldsymbol{X},\boldsymbol{U}_R}\left[\sum_{j=1}^J f_j(X_j)|\boldsymbol{U}_F,\boldsymbol{\sigma}_R\right]|\boldsymbol{\sigma}_F\right\}
 &=\sum_{j=L+1}^J E_{\boldsymbol{X}}\left\{E_{\boldsymbol{U}_R}\left[f_j^2(X_j)|\boldsymbol{X},\boldsymbol{\sigma}_R\right]\right\}\\
 &+E_{\boldsymbol{U}_F}\left\{E_{\boldsymbol{X}}\left\{\left[\sum_{j=1}^L f_j(X_j)\right]^2|\boldsymbol{U}_F\right\}|\boldsymbol{\sigma}_F\right\}.
\end{align*}
Inverting the order of expectation, we get:
\begin{align*}
E_{\boldsymbol{U}_F}\left\{Var_{\boldsymbol{X},\boldsymbol{U}_R}\left[\sum_{j=1}^J f_j(X_j)|\boldsymbol{U}_F,\boldsymbol{\sigma}_R\right]|\boldsymbol{\sigma}_F\right\}
 &=\sum_{j=L+1}^J E_{\boldsymbol{X}}\left\{E_{\boldsymbol{U}_R}\left[f_j^2(X_j)|\boldsymbol{X},\boldsymbol{\sigma}_R\right]\right\}\\
 &+E_{\boldsymbol{X}}\left\{E_{\boldsymbol{U}_F}\left\{\left[\sum_{j=1}^L f_j(X_j)\right]^2|\boldsymbol{\sigma}_F\right\}\right\}\\
  &=\sum_{j=L+1}^J E_{\boldsymbol{X}}\left\{E_{\boldsymbol{U}_R}\left[f_j^2(X_j)|\boldsymbol{X},\boldsymbol{\sigma}_R\right]\right\}\\
 &+\sum_{j=1}^L E_{\boldsymbol{X}}\left\{E_{\boldsymbol{U}_F}\left[ f_j^2(X_j)\right]|\boldsymbol{\sigma}_F\right\}\\
 &=\sum_{j=1}^J E_{\boldsymbol{X}}\left\{E_{\boldsymbol{U}_F,\boldsymbol{U}_R}\left[ f_j^2(X_j)|\boldsymbol{\sigma}_F,\boldsymbol{\sigma}_R\right]\right\}\\
 &=\sum_{j=1}^J E_{\boldsymbol{X}}\left\{Var_{\boldsymbol{u}_1,...,\boldsymbol{u}_J}\left[ f_j(X_j)|\boldsymbol{\sigma}\right]\right\}\\
  &=\sum_{j=1}^J \sigma^2_j E_{X_j}\left[\boldsymbol{D}^T_j(X_j)\boldsymbol{Q}^*_j\boldsymbol{D}_j(X_j)\right]
\end{align*}
If scaling has been applied as in Proposition \ref{prop:standardization}, then we know that:
\begin{gather*}
\sum_{j=1}^J\sigma^2_j E_{X_j}\left[\boldsymbol{D}^T_j(X_j)\boldsymbol{Q}^*_j\boldsymbol{D}_j(X_j)\right]=\sum_{j=1}^J\sigma^2_j
\end{gather*}
which completes the proof.

\section{Proof of Example \ref{ex:2}}\label{sm:example_2}
Equation 2.29 of \citet{RH} can be used to find that the covariance matrix for an originally i.i.d. effect under constraint $\boldsymbol{a}^T\boldsymbol{u}=\boldsymbol{0}$ becomes equal to:
\begin{align*}
\boldsymbol{Q}^*&= \boldsymbol{I}-\boldsymbol{Ia}(\boldsymbol{a}^T\boldsymbol{Ia})^{-1}\boldsymbol{a}^T\boldsymbol{I}\\
&= \boldsymbol{I}-\boldsymbol{a}(\boldsymbol{a}^T\boldsymbol{a})^{-1}\boldsymbol{a}^T\\
&= \boldsymbol{I}-\dfrac{\boldsymbol{aa}^T}{\boldsymbol{a}^T\boldsymbol{a}}.
\end{align*}
In order for $ \boldsymbol{a}^T\boldsymbol{u}=\boldsymbol{0}\implies E_X[f(X)])=0$, $\boldsymbol{a}$ must be equal to $[p_1,...,p_K]^T$:
  \begin{align*}
    \boldsymbol{a}^T\boldsymbol{u}=\boldsymbol{0}&= E_X[f(X)]\\
    &= \sum_{k=1}^K p_k f(k)\\
    &= \sum_{k=1}^K p_k \left[\sum_{j=1}^K \mathbb{I}(k=j) u_j\right]\\
    &= \sum_{k=1}^K p_k  u_k\\
    &= \begin{bmatrix}
        p_1 & p_2 & ... & p_K
    \end{bmatrix} \boldsymbol{u}
\end{align*}

Finally, $C$ can be found applying Proposition \ref{prop:standardization}, knowing $\boldsymbol{Q}^*$ and $\boldsymbol{a}$:
\begin{align*}
C&=\sum_{k=1}^K p_k\boldsymbol{D}(k)\left[\boldsymbol{I}-\dfrac{\boldsymbol{aa}^T}{\boldsymbol{a}^T\boldsymbol{a}}\right]\boldsymbol{D}^T(k)\\
&=1-\sum_{k=1}^K p_k\boldsymbol{D}^T(k)\left[\dfrac{\boldsymbol{aa}^T}{\boldsymbol{a}^T\boldsymbol{a}}\right]\boldsymbol{D}(k)\\
&=1-\dfrac{1}{\boldsymbol{a}^T\boldsymbol{a}}\sum_{k=1}^K p_k\boldsymbol{D}^T(k)\boldsymbol{aa}^T\boldsymbol{D}(k)\\
&=1-\dfrac{1}{\sum_{k=1}^K p_k^2}\sum_{k=1}^K p_k\left[\sum_{j=1}^K \mathbb{I}(j=k)p_j\right]^2\\
&=1-\dfrac{1}{\sum_{k=1}^K p_k^2}\sum_{k=1}^K p_k^3\\
&=1-\dfrac{\sum_{k=1}^K p_k^3}{\sum_{k=1}^K p_k^2}\\
\end{align*}

\section{Proof of Eq. (\ref{eq:null_trend})}\label{sm:IGMRF_discrete}
Remembering the definition of $\boldsymbol{S}_{(d-1)}^T\boldsymbol{u}=\boldsymbol{0}$ from Definition \ref{def:IGMRF}, it can be proven that $\boldsymbol{S}_{(d-1)}^T\boldsymbol{u}=\boldsymbol{0}$ implies $E_X[X^m f_r(X)]=0,m=0,...,d-1$ when $f_r(X)$ is defined as in Section \ref{sec:disc_IGMRFs} and $X\sim \text{Unif}[1,K]$. First, $\boldsymbol{S}_{(d-1)}^T\boldsymbol{u}$ can be written explicitly as:
\begin{align*}
\boldsymbol{S}_{(d-1)}^T\boldsymbol{u}&=\begin{bmatrix}
\sum_{k=1}^K k^0  u_k\\
\sum_{k=1}^K k^1 u_k\\
...\\
\sum_{k=1}^K k^{d-1} u_k
\end{bmatrix}
\end{align*}
Hence, if  $\boldsymbol{S}_{(d-1)}^T\boldsymbol{u}=\boldsymbol{0}$, then $\sum_{k=1}^K k^{m} u_k=0$ for $m=0,...,d-1$. 
Secondly, we can also rewrite $E_X[X^m f_r(X)]=0$ explicitly:
\begin{align*}
E_X[X^m f_r(X)]=\dfrac{1}{K}\sum_{k=1}^K k^m u_k
\end{align*}
Since $\sum_{k=1}^K k^{m} u_k=0$ implies that $K^{-1}\sum_{k=1}^K k^{m} u_k=0$  for $m=0,...,d-1$, it is therefore proven that in this case $\boldsymbol{S}_{(d-1)}^T\boldsymbol{u}=\boldsymbol{0}$ implies $E_X[X^m f_r(X)]=0$ for $m=0,...,d-1$.

\section[sectionsm]{Design of $f_t(X)$ from Section \ref{sec:disc_IGMRFs}}\label{sm:f_t}
For $d>2$, $f_t(X)$ must represent a polynomial trend of degree $d-1$. In order to obtain a simple interpretation for the variance parameters associated with $f_t(X)$, we propose to introduce the effect as:
\begin{equation}\label{eq:f_t_def}
\begin{aligned}
    f_t(X)&=\sum_{m=1}^{d-1}f_{t_m}(X)\\
    &=\sum_{m=1}^{d-1}h_m(X)\beta_m
\end{aligned}
\end{equation}
where $f_{t_1}(X),...,f_{t_{d-1}}(X)$ are 
$d-1$ separate effects with polynomial basis functions $h_1(X),...,h_{d-1}(X)$, designed such that each of them respectively represents the linear, quadratic, cubic, etc. contribution to the effect of $X$. Each of the $\beta_1,...,\beta_{d-1}$ coefficients is then assigned a separate variance parameter:
\begin{gather*}
 \beta_m \sim N(0,\sigma^2_m),\;\;m=1,...,d-1
\end{gather*}
so that each $\sigma^2_m$ can measure (after standardization) the $m^{th}$-degree polynomial contribution to the variance.

The intuitive interpretation of each of the  $\sigma^2_m$ as exclusively the contribution of their corresponding polynomial degree is achieved by designing each $h_m(X)$ as a polynomial function of degree $m$ :
\begin{gather}\label{eq:h_func_def}
h_m(x)=\sum_{l=0}^m a_{m,l} x^l
\end{gather}
whose $a_0,...,a_m$ must be constrained such that any polynomial trend of degree $m-1$ is removed, i.e.:
\begin{align}\label{eq:h_func_constr}
    E_X[ X^l  h_m(x)]=0 \quad \quad\forall \; l=0,...,m-1.
\end{align}
Under these constraints, we can find for example $h_1(X)=X-E[X]$ and $h_2(X)=X^2-E[X^2]-\dfrac{Cov[X,X^2](X-E[X])}{Var[X]}$.
\section{Proof of Eq. (\ref{eq:IGMRF_var})}
Consider $f_{\text{new}}(X)=f_t(X)+f_r(X)$ where $f_t(X)$ is defined as Eq. (\ref{eq:f_t_def}) from Section \ref{sm:IGMRF_discrete} of the Supplementary Material, and $f_r(X)$ is the IGMRF effect from Eq. (\ref{eq:discrete_IGMRF}) of the main paper under null space constraints. The variance of $f_{\text{new}}(X)$ with respect to both $X$ and the coefficients $\beta_1,..,\beta_{d-1},\boldsymbol{u}$ is equal to:
\begin{align*}
Var_{X,\beta_1,...,\beta_{d-1},\boldsymbol{u}}[f_{\text{new}}(X)]&=Var_{X,\beta_1,...,\beta_{d-1},\boldsymbol{u}}\left[\sum_{m=1}^{d-1}f_{t_m}(X)+f_r(X)\right]\\
&=E_{X}\left\{Var_{\beta_1,...,\beta_{d-1},\boldsymbol{u}}\left[\sum_{m=1}^{d-1}f_{t_m}(X)+f_r(X)\right]\right\}\\
&=\sum_{m=1}^{d-1}E_{X}\{Var_{\beta_m}[f_{t_m}(X)]\}\\
&+E_{X}\{Var_{\boldsymbol{u}}[f_{r}(X)]\}\\
&+2\sum_{l<m}E_{X}\{Cov_{\beta_l,\beta_m}[f_{t_l}(X),f_{t_m}(X)]\}\\
&+2\sum_{m=1}^{d-1}E_{X}\{Cov_{\beta_m,\boldsymbol{u}}[f_{t_m}(X),f_{r}(X)]\}
\end{align*}
If all the effects in $f_{\text{new}}(X)$ have been appropriately standardized (Proposition \ref{prop:standardization}), then the variance simplifies to:
\begin{align*}
Var_{X,\beta_1,...,\beta_{d-1},\boldsymbol{u}}[f_{\text{new}}(X)]&=\sum_{m=1}^{d-1}\sigma^2_m+\sigma^2_r\\
&+2\sum_{l<m}E_{X}\{Cov_{\beta_l,\beta_m}[f_{t_l}(X),f_{t_m}(X)]\}
\\
&+2\sum_{m=1}^{d-1}E_{X}\{Cov_{\beta_m,\boldsymbol{u}}[f_{t_m}(X),f_{r}(X)]\}
\end{align*}
To show that all the covariance terms are null, we can use the properties of the distributions chosen on the parameters, specifically null mean and mutual independence assumption between $\beta_1,...,\beta_{d-1},\boldsymbol{u}$. For $l=1,...,d-2$ and $m> l$:
\begin{align*}
E_{X}\{Cov_{\beta_l,\beta_m}[f_{t_l}(X),f_{t_m}(X)]\}&=E_{X}\{E_{\beta_l,\beta_m}[f_{t_l}(X) f_{t_m}(X)]\}\\
&-E_{X}\{E_{\beta_l}[f_{t_l}(X)] E_{\beta_m}[f_{t_m}(X)]\}\\
&=E_{X}\{E_{\beta_l,\beta_m}[h_l(X)\beta_l h_m(X)\beta_m]\}\\
&-E_{X}\{E_{\beta_l}[h_l(X)\beta_l] E_{\beta_m}[h_m(X)\beta_m]\}\}\\
&=0
\end{align*}
Additionally, for $m=1,...,d-1$:
\begin{align*}
E_{X}\{Cov_{\beta_m,\boldsymbol{u}}[f_{t_m}(X),f_{r}(X)]\}&=E_{X}\{E_{\beta_m,\boldsymbol{u}}[f_{t_l}(X) f_{r}(X)]\}\\
&-E_{X}\{E_{\beta_m}[f_{t_m}(X)] E_{\boldsymbol{u}}[f_{r}(X)]\}\\
&=E_{X}\{E_{\beta_m,\boldsymbol{u}}[h_m(X)\beta_m  \mathbf{D}^T(X)\mathbf{u}]\}\\
&-E_{X}\{E_{\beta_m}[h_m(X)\beta_m ] E_{\boldsymbol{u}}[\mathbf{D}^T(X)\mathbf{u}]\}\\
&=0 
\end{align*}
Hence, we can finally write the variance of $f_{\text{new}}(X)$ as:
\begin{align*}
Var_{X,\beta_1,...,\beta_{d-1},\boldsymbol{u}}[f_{\text{new}}(X)]&=\sum_{m=l}^{d-1}\sigma^2_m+\sigma^2_r
\end{align*}

\section{Proof of Example \ref{ex:4}}\label{sm:example_4}
The first step of the Q modification consists in explicitly deriving the entries of the desired null space, i.e. $\widetilde{\boldsymbol{S}}$ as defined in Eq. (\ref{eq:general_S_tilde}). In this case, the matrix has two columns $\widetilde{\boldsymbol{S}}=[\widetilde{\boldsymbol{s}}_0\quad
          \widetilde{\boldsymbol{s}}_1]$ where:
\begin{align*}
     \widetilde{\boldsymbol{s}}_0&=E_X[\boldsymbol{B}(X)] \\
\widetilde{\boldsymbol{s}}_1&=E_X[X \boldsymbol{B}(X)] 
\end{align*}
In order to explicitly derive  $\widetilde{\boldsymbol{S}}$ is necessary to first derive explicit expressions for $\boldsymbol{B}(x)$. In order to derive analytically the B-spline basis, it is first necessary to make explicit the degree $d$ and the number of basis functions $K$ so that
a B-spline can be denoted by $\boldsymbol{B}^{(d)}_K(X)$. Hence, the cubic B-spline is actually $\boldsymbol{B}(X)=\boldsymbol{B}^{(3)}_K(X)$.  \citet{EM} defined analytically equally spaced B-splines with the recursion formula in Equation 45:
\begin{equation}
\begin{aligned}\label{eq:Bspline_formula}
 B_k^{(0)}(X)&=\mathbb{I}\left[\dfrac{k-1}{K-D}<\dfrac{X-m}{M-m}<\dfrac{k}{K-D}\right]\\
    B_k^{(d)}(X)&=
    \begin{cases}
     \frac{1}{d}\left[d+\frac{X-m}{M-m}(K-D)-k+1\right]B_{k-1}^{(d-1)}(X)+\\
     \frac{1}{d}\left[k-\frac{X-m}{M-m}(K-D)\right]B_{k}^{(d-1)}(X),\quad 1\leq k\leq K-D+d\\
     \;\\
     0,\quad \text{otherwise}
    \end{cases}
\end{aligned}
\end{equation}
From the recursive formula of Eq. (\ref{eq:Bspline_formula}), it can be found that the elements of $\boldsymbol{B}(x)=[B_1(x),...,B_K(x)]$ are:
  \begin{align*}
B_k(x)= \Bigg[
  &\mathbb{I}(k-1<\hat{x}(K-3)<k)\cdot g_1\left(\hat{x}(K-3)-(k-1)\right)+\\
  &\mathbb{I}(k-2<\hat{x}(K-3)<k-1)\cdot g_2\left(\hat{x}(K-3)-(k-2)\right)+\\
  &\mathbb{I}(k-3<\hat{x}(K-3)<k-2)\cdot g_3\left(\hat{x}(K-3)-(k-3)\right)+\\
  &\mathbb{I}(k-4<\hat{x}(K-3)<k-3)\cdot g_4\left(\hat{x}(K-3)-(k-4)\right)    
  \Bigg    ]\mathbb{I}(0<\hat{x}<1)
\end{align*}
where $\hat{x}=\dfrac{x-m}{M-m}\in[0,1]$ and:
\begin{align*}
g_1(y)&= \dfrac{1}{2}\left[-\dfrac{{y}^3}{3}+{y}^2-y+\dfrac{1}{3}\right]\\
g_2(y)&= \dfrac{y^3}{2}-{y}^2+\dfrac{2}{3}\\
g_3(y)&= \dfrac{1}{2}\left[-{y}^3+{y}^2+y+\dfrac{1}{3}\right]\\
g_4(y)&=\dfrac{{y}^3}{6} 
\end{align*}
Noting that $\boldsymbol{B}(x)$ is in fact only a function of the normalized version $\hat{x}$:
\begin{align*}
    \widetilde{\boldsymbol{s}}_0&=
     (M-m)\int_{0}^1 \boldsymbol{B}(\hat{x}) \pi((M-m)\hat{x}+m)\diff \hat{x} \\
      \widetilde{\boldsymbol{s}}_1&=(M-m) m  \widetilde{\boldsymbol{s}}_0+
     (M-m)^2\int_{0}^1 \hat{x} \boldsymbol{B}(\hat{x}) \pi((M-m)\hat{x}+m)\diff \hat{x} 
\end{align*}
If $X \sim \text{Unif}(m,M)$ so that $\pi(x)=\dfrac{I(m<x<M)}{M-m}$:
\begin{align*}
  \widetilde{\boldsymbol{s}}_0
   &= \int_{0}^1 \boldsymbol{B}(\hat{x})\diff \hat{x}\\
   &= \dfrac{1}{K-3} \left[\dfrac{1}{24},\dfrac{1}{2},\dfrac{23}{24},1,....,1,\dfrac{23}{24},\dfrac{1}{2},\dfrac{1}{24}\right]^T ,\;\;K\geq7\\
   \widetilde{\boldsymbol{s}}_1
&=m \widetilde{\boldsymbol{s}}_0+(M-m)\left[\int_{0}^1 \hat{x}
\boldsymbol{B}(\hat{x})\diff \hat{x}\right] \\
&=m \widetilde{\boldsymbol{s}}_0+(M-m)\boldsymbol{v}
\end{align*}
where $\boldsymbol{v}=[v_1,...,v_K]^T$ for $K\geq 7$:
\begin{align*}
v_1&=\dfrac{1}{(K-3)^2}\cdot \dfrac{1}{120}\\
v_2&=\dfrac{1}{(K-3)^2} \cdot\dfrac{7}{30}\\
v_3&=\dfrac{1}{(K-3)^2} \cdot\dfrac{121}{120}\\
v_k&=\dfrac{k-2}{(K-3)^2} \;\;\;k=4,...,K-3\\
v_{K-2}&= \dfrac{1}{K-3}\cdot\dfrac{23}{24}-\dfrac{1}{(K-3)^2} \cdot\dfrac{121}{120}\\
v_{K-1}&=\dfrac{1}{K-3}\cdot\dfrac{1}{2}-\dfrac{1}{(K-3)^2} \cdot\dfrac{7}{30}\\
v_K&=\dfrac{1}{K-3}\cdot\dfrac{1}{24}-\dfrac{1}{(K-3)^2}\cdot \dfrac{1}{120}
\end{align*}
For $K=6$:
\begin{align*}
\begin{matrix}
    \widetilde{\boldsymbol{s}}_0=\left[\dfrac{1}{72},\dfrac{1}{6},\dfrac{23}{72},\dfrac{23}{72},\dfrac{1}{6},\dfrac{1}{72}\right]^T & 
    \boldsymbol{v}=\left[\dfrac{1}{1080},\dfrac{7}{270},\dfrac{121}{1080},\dfrac{28}{135},\dfrac{19}{135},\dfrac{7}{540}\right]^T
\end{matrix}
\end{align*}
For $K=5$:
\begin{align*}
\begin{matrix}
\widetilde{\boldsymbol{s}}_0=\left[\dfrac{1}{48},\dfrac{1}{4},\dfrac{11}{24},\dfrac{1}{4},\dfrac{1}{48}\right]^T & 
    \boldsymbol{v}=\left[\dfrac{1}{480},\dfrac{7}{120},\dfrac{11}{48},\dfrac{23}{120},\dfrac{3}{160}\right]^T
    \end{matrix}
\end{align*}
For $K=4$:
\begin{align*}
    \begin{matrix}
\widetilde{\boldsymbol{s}}_0=\left[\dfrac{1}{24},\dfrac{11}{24},\dfrac{11}{24},\dfrac{1}{24}\right]^T & 
         \boldsymbol{v}=\left[\dfrac{1}{120},\dfrac{11}{60},\dfrac{11}{40},\dfrac{1}{30}\right]^T
    \end{matrix} 
\end{align*}
Now that the entries of the desired null space $\widetilde{\boldsymbol{S}}$ are available, the new precision matrix $\widetilde{\boldsymbol{Q}}$ can be designed. If it is specified as $\widetilde{\boldsymbol{Q}}=(\boldsymbol{\Lambda}\widetilde{\boldsymbol{R}}^*\boldsymbol{\Lambda})^*$ as in Eq. (\ref{eq:decomposition}), and if $\widetilde{\boldsymbol{R}}=\widetilde{\boldsymbol{G}}-\widetilde{\boldsymbol{W}}$ where $\widetilde{\boldsymbol{W}}$ and $\widetilde{\boldsymbol{G}}$ are defined respectively as in Equations \ref{eq:new_W_ex_5} and \ref{eq:new_G_ex_5}, then $\widetilde{\boldsymbol{Q}}\widetilde{\boldsymbol{S}}=\boldsymbol{0}$ if $\widetilde{\boldsymbol{G}}\boldsymbol{\Lambda}\widetilde{\boldsymbol{S}}-\widetilde{\boldsymbol{W}}\boldsymbol{\Lambda}\widetilde{\boldsymbol{S}}=\boldsymbol{0}$. Making this expression explicit, we get:
\begin{eqnarray*}
\widetilde{\boldsymbol{G}}\boldsymbol{\Lambda}\widetilde{\boldsymbol{S}}-\widetilde{\boldsymbol{W}}\boldsymbol{\Lambda}\widetilde{\boldsymbol{S}}&=&\begin{bmatrix}
        \widetilde{G}_{1,1} \lambda_1\widetilde{S}_{1,0} & \widetilde{G}_{1,1} \lambda_1\widetilde{S}_{1,1}\\
        \widetilde{G}_{2,2} 
        \lambda_2\widetilde{S}_{2,0} & \widetilde{G}_{2,2} 
        \lambda_2\widetilde{S}_{2,1}\\
         ... & ...\\
         \widetilde{G}_{K,K}
         \lambda_K\widetilde{S}_{K,0} & \widetilde{G}_{K,K}
         \lambda_K\widetilde{S}_{K,1}
    \end{bmatrix}\\
    &-&
    \begin{bmatrix}
        \sum_{l=1}^K\widetilde{W}_{1,l} 
        \lambda_l\widetilde{S}_{l,0} & \sum_{l=1}^K\widetilde{W}_{1,l} 
         \lambda_l
       \widetilde{S}_{l,1}\\
\sum_{l=1}^K\widetilde{W}_{2,l} 
 \lambda_l
      \widetilde{S}_{l,0} &  \sum_{l=1}^K\widetilde{W}_{2,l} 
       \lambda_l
       \widetilde{S}_{l,1}\\
         ... & ...\\
      \sum_{l=1}^K\widetilde{W}_{K,l} 
       \lambda_l
      \widetilde{S}_{l,0} & \sum_{l=1}^K\widetilde{W}_{K,l} 
       \lambda_l\widetilde{S}_{l,1}
    \end{bmatrix}
\end{eqnarray*}
Therefore, $\widetilde{\boldsymbol{Q}}\widetilde{\boldsymbol{S}}=\boldsymbol{0}$ if:
\begin{gather*}
   \begin{bmatrix}
        \widetilde{G}_{k,k} \lambda_k \widetilde{S}_{k,0} & \widetilde{G}_{k,k} \lambda_k\widetilde{S}_{k,1}
    \end{bmatrix}=
    \begin{bmatrix}
\sum_{l=1}^K\widetilde{W}_{k,l}\lambda_l \widetilde{S}_{l,0} & \sum_{l=1}^K\widetilde{W}_{k,l}\lambda_l \widetilde{S}_{l,1}
    \end{bmatrix}
\end{gather*}
for $k=1,...,K$. Replacing the entries of $\widetilde{\boldsymbol{G}}$ with their definition from Eq. (\ref{eq:new_G_ex_5}), we obtain:
\begin{gather*}
   \begin{bmatrix}
     \sum_{l=1}^K  \widetilde{W}_{k,l}  \lambda_l\widetilde{S}_{l,0} & \frac{\widetilde{S}_{k,1}}{\widetilde{S}_{k,0}}\sum_{l=1}^K  \widetilde{W}_{k,l} \lambda_l\widetilde{S}_{l,0}
    \end{bmatrix}=
    \begin{bmatrix}
        \sum_{l=1}^K\widetilde{W}_{k,l}\lambda_l\widetilde{S}_{l,0} & \sum_{l=1}^K\widetilde{W}_{k,l} \lambda_l\widetilde{S}_{l,1}
    \end{bmatrix}
\end{gather*}
Hence, since the first elements of both vectors are equal, it is only necessary to verify that:
\begin{gather*}
    \dfrac{\widetilde{S}_{k,1}}{\widetilde{S}_{k,0}}\sum_{l=1}^K  \widetilde{W}_{k,l} \lambda_l\widetilde{S}_{l,0}-\sum_{l=1}^K  \widetilde{W}_{k,l} \lambda_l\widetilde{S}_{l,1}=0\\
   \Downarrow\\
   \sum_{l=1}^K\widetilde{W}_{k,l} \lambda_l (\widetilde{S}_{l,0} \widetilde{S}_{k,1}-\widetilde{S}_{l,1} \widetilde{S}_{k,0})=0
\end{gather*}
for $k=1,...,K$. Replacing the entries of $\widetilde{\boldsymbol{W}}$  with their definition from Eq. (\ref{eq:new_W_ex_5}), these $K$ conditions are transformed into:
\begin{gather*}
\lambda_k^{-1} \sum_{l=1}^K (k-l) W_{k,l}=0 \quad k=1,...,K
\end{gather*}
which is true for all $k$ for the definition of $\boldsymbol{W}$ given in Example \ref{ex:4}.

There might be cases in which it is desirable to specify P-splines with a first-order random walk process on the coefficients, i.e. $\boldsymbol{u}|\sigma^2_r\sim N(\boldsymbol{0},\sigma^2_r \boldsymbol{Q}_{\text{RW1}}^*)$ where $\boldsymbol{Q}_{\text{RW1}}$ has been defined in Eq. (\ref{eq:prelim_RW1}). In this case, the new null space must be equal to:
\begin{gather*}
  \underset{K\times 1}{\widetilde{\boldsymbol{S}}}=\begin{bmatrix}
       \widetilde{S}_1\\
        \widetilde{S}_2\\
        ...\\
         \widetilde{S}_{K}
    \end{bmatrix}=
    \begin{bmatrix}
        E_X[\boldsymbol{B}(X)]
    \end{bmatrix}
\end{gather*}
A valid solution for the modified precision matrix comes from designing $\widetilde{\boldsymbol{R}}=\widetilde{\boldsymbol{G}}-\widetilde{\boldsymbol{W}}$ where:
\begin{align*}
     \widetilde{W}_{k,l}&=\dfrac{{W}_{k,l}}{ \lambda_k \widetilde{S}_k \lambda_l\widetilde{S}_l}\\
      \widetilde{G}_{k,l}&=\dfrac{G_{k,l}}{\lambda_k^2 \widetilde{S}_k^2}
\end{align*}
where $\boldsymbol{W}=\boldsymbol{G}-\boldsymbol{Q}_{\text{RW1}}$ and $\boldsymbol{G}=\text{diag}(\text{diag}(\boldsymbol{Q}_{\text{RW1}}))$. Then, $\widetilde{\boldsymbol{Q}}\widetilde{\boldsymbol{S}}=\boldsymbol{0}$ if $\widetilde{\boldsymbol{G}}\boldsymbol{\Lambda}\widetilde{\boldsymbol{S}}-\widetilde{\boldsymbol{W}}\boldsymbol{\Lambda}\widetilde{\boldsymbol{S}}=\boldsymbol{0}$. We can find that:
\begin{align*}
    \widetilde{\boldsymbol{G}}\boldsymbol{\Lambda}\widetilde{\boldsymbol{S}}-\widetilde{\boldsymbol{W}}\boldsymbol{\Lambda}\widetilde{\boldsymbol{S}}
         &= \begin{bmatrix}
         \widetilde{G}_{1,1}\lambda_1\widetilde{S}_1\\
        ...\\
        \widetilde{G}_{K,K} \lambda_{K}\widetilde{S}_{K}
    \end{bmatrix}-\begin{bmatrix}
       \sum_{l=1}^{K}\widetilde{W}_{1,l}\lambda_l\widetilde{S}_l\\
        ...\\
         \sum_{l=1}^{K}\widetilde{W}_{K,l}\lambda_l\widetilde{S}_l
    \end{bmatrix}
\end{align*}
Replacing $\widetilde{\boldsymbol{G}}$ and $\widetilde{\boldsymbol{W}}$ with their definitions in terms of $\boldsymbol{G}$ and $\boldsymbol{W}$, it is found that:
\begin{align*}
    \widetilde{\boldsymbol{G}}\boldsymbol{\Lambda}\widetilde{\boldsymbol{S}}-\widetilde{\boldsymbol{W}}\boldsymbol{\Lambda}\widetilde{\boldsymbol{S}}
         &= \begin{bmatrix}
         \dfrac{G_{1,1}}{\lambda_1\widetilde{S}_1}\\
        ...\\
        \dfrac{G_{K,K}}{\lambda_{K}\widetilde{S}_{K}}
    \end{bmatrix}-\begin{bmatrix}
       \dfrac{ \sum_{l=1}^{K}W_{1,l}}{\lambda_1\widetilde{S}_1}\\
        ...\\
         \dfrac{\sum_{l=1}^{K}W_{K,l}}{\lambda_{K}\widetilde{S}_{K}}
    \end{bmatrix}\\
    &= \begin{bmatrix}
         \dfrac{G_{1,1}}{\lambda_1\widetilde{S}_1}\\
        ...\\
        \dfrac{G_{K,K}}{\lambda_{K}\widetilde{S}_{K}}
    \end{bmatrix}-\begin{bmatrix}
         \dfrac{G_{1,1}}{\lambda_1\widetilde{S}_1}\\
        ...\\
        \dfrac{G_{K,K}}{\lambda_{K}\widetilde{S}_{K}}
    \end{bmatrix}=\boldsymbol{0}
\end{align*}
In this case, the order of the IGMRF is $d=1$ so that the trend effect is not necessary and the P-spline model can be written simply as $f(X)=f_r(X)$:
\begin{equation*}
\begin{aligned}
    f(X)&=\boldsymbol{B}^{T}(X)\boldsymbol{u}\\
\boldsymbol{u}|\sigma^2&\sim N(\boldsymbol{0},\sigma^2 \widetilde{\boldsymbol{Q}}^*)\text{ subject to } \widetilde{\boldsymbol{S}}^T\boldsymbol{u}=0
\end{aligned}
\end{equation*}

\section{Scaling constants table from Section \ref{sec:IGMRFs}}
\begin{table}[H]
    \centering
    \begin{tabular}{r|rrrr}
\toprule
$K$ & RW1 & RW2 & P-spline (RW1)& P-spline (RW2)\\
        \midrule
      5 & 0.800 & 0.300 & 0.152 & 0.012 \\ 
 6 & 0.972 & 0.521 & 0.289 & 0.045 \\ 
 7 & 1.143 & 0.827 & 0.440 & 0.123 \\ 
 8 & 1.313 & 1.232 & 0.596 & 0.266 \\ 
 9 & 1.481 & 1.752 & 0.756 & 0.496 \\ 
 10 & 1.650 & 2.400 & 0.917 & 0.835 \\ 
 12 & 1.986 & 4.139 & 1.244 & 1.924 \\ 
 15 & 2.489 & 8.068 & 1.738 & 4.695 \\ 
 20 & 3.325 & 19.093 & 2.566 & 13.328 \\ 
 25 & 4.160 & 37.260 & 3.397 & 28.438 \\ 
 30 & 4.994 & 64.356 & 4.229 & 51.693 \\ 
 40 & 6.662 & 152.475 & 5.895 & 129.476 \\ 
 50 & 8.330 & 297.737 & 7.562 & 259.966 \\ 
 100 & 16.665 & 2381.19 & 15.901 & 2164.456 \\ 
        \bottomrule
       \end{tabular}
       \caption{Scaling constants for $f_r(X)$ for Examples \ref{ex:3} and \ref{ex:4} for different values of $K$.}\label{tab:scaling_constants}
\end{table}

\section{Proof of Eq. (\ref{eq:varphi_induced})}\label{sm:sim_prior}
For a given prior on $\sigma^2, \sigma^2_\epsilon$, the implied prior on $V$ and $\omega$ is:
\begin{eqnarray*}
    \pi(V,\omega)&=&\pi_{\sigma^2}(V\cdot \omega)\cdot\pi_{\sigma^2_\epsilon}(V-V\cdot \omega)\cdot \bigg | \det 
    \begin{bmatrix}
        \frac{\diff V\omega}{\diff \omega} & \frac{\diff V\omega}{\diff V}\\
        \frac{\diff V(1-\omega)}{\diff \omega} & \frac{\diff V(1-\omega)}{\diff V}
    \end{bmatrix}
    \Bigg |\\
    &=&\pi_{\sigma^2}(V\cdot \omega)\cdot\pi_{\sigma^2_\epsilon}(V-V\cdot \omega)\cdot V
\end{eqnarray*}

The marginal of $\varphi$ implied by a prior specification on $V,\omega$ can be found in two steps. First, the marginal of $\omega$ is found marginalizing out $V$: 
\begin{gather*}
   \pi(\omega)=\int_0^{\infty} \pi(V,\omega) \diff V
\end{gather*}
Secondly, the marginal of $\varphi$ is found through a change of variable formula using the transformation $\omega=\dfrac{\varphi }{\varphi +C-\varphi C}$:
\begin{gather*}
   \pi(\varphi )=\pi_\omega\left(\dfrac{\varphi }{\varphi +C-\varphi C}\right) \dfrac{C}{\left[\varphi +C-\varphi C\right]^2}
\end{gather*}
For the 3 prior specifications, we can derive the implied prior on $\varphi $.
\begin{enumerate}[label=(\alph*)]
    \item \textbf{IG priors}: $\sigma^2,\sigma^2_\epsilon \overset{iid}{\sim} \text{IG}(1,\beta)$ 
First, we derive the implied prior on $V,\omega$.
    \begin{eqnarray*}
        \pi(V,\omega)&=&\beta^2 V^{-4} \omega^{-2}(1-\omega)^{-2}\exp\left[-\frac{\beta}{V}\left(\frac{1}{\omega}+\frac{1}{1-\omega}\right)\right] V\\
&=&\beta^2 V^{-3} \omega^{-2}(1-\omega)^{-2}\exp\left[-\frac{\beta}{V}\left(\frac{1}{\omega}+\frac{1}{1-\omega}\right)\right]
 \end{eqnarray*}
Secondly, we marginalize out $V$.
    \begin{eqnarray*}
\pi(\omega)&=& \beta^2  \omega^{-2}(1-\omega)^{-2}\int_0^{\infty}\exp\left[-\frac{\beta}{V}\left(\frac{1}{\omega}+\frac{1}{1-\omega}\right)\right]V^{-3} \diff V\\
&=&\left[\frac{\beta}{\omega(1-\omega)}\right]^{2}\left[\frac{\beta}{\omega(1-\omega)}\right]^{-2}\\
&=&1
\end{eqnarray*}
Finally, we find the implied prior on $\varphi$.
    \begin{eqnarray*}
\pi(\varphi )&=&\dfrac{C}{\left[\varphi +C-\varphi C\right]^{2}}
    \end{eqnarray*}
    \item \textbf{PC priors}: $\sigma^2,\sigma^2_\epsilon  \overset{iid}{\sim}\text{PC}_{0}(\delta)$ 
    \begin{eqnarray*}
\pi(V,\omega)&=&\delta^2 \exp(-\delta\sqrt{V\omega})\exp(-\delta\sqrt{V-V\omega})\frac{1}{4V\sqrt{\omega(1-\omega)}} V\\
&=&\frac{\delta^2}{4\sqrt{\omega(1-\omega)}} \exp[-\delta\sqrt{V}(\sqrt{\omega}+\sqrt{1-\omega})] \end{eqnarray*}
Secondly, we marginalize out $V$.
    \begin{eqnarray*}
\pi(\omega)&=&\frac{\delta^2}{4\sqrt{\omega(1-\omega)}} \int_{0}^{\infty}\exp[-\delta\sqrt{V}(\sqrt{\omega}+\sqrt{1-\omega})]\diff V\\
&=&
\frac{\delta^2}{4\sqrt{\omega(1-\omega)}}\frac{2}{\delta^2(\sqrt{\omega}+\sqrt{1-\omega})^2}\\
&=&\left[2\sqrt{\omega(1-\omega)}(\sqrt{\omega}+\sqrt{1-\omega})^2\right]^{-1}
\end{eqnarray*}
Finally, we find the implied prior on $\varphi$.
    \begin{eqnarray*}
        \pi(\varphi )&=&\left[2\sqrt{C\varphi (1-\varphi )}\left(\sqrt{\frac{\varphi }{C}}+\sqrt{1-\varphi }\right)^2\right]^{-1}
    \end{eqnarray*}
    \item \textbf{VP prior}: $\omega\sim $Beta$(\alpha,\beta)$ (generalization of the Uniform case)
    
    In this case, the prior is simply found through the change of variable formula: 
    \begin{gather*}
    \pi(\varphi )=\dfrac{\Gamma(\alpha+\beta)}{\Gamma(\alpha)\Gamma(\beta)} \dfrac{C (\varphi )^{\alpha-1} (C-\varphi C)^{\beta-1}}{[\varphi +C-\varphi C]^{\alpha+\beta}}
    \end{gather*}
\end{enumerate}

\section{Additional simulation results from Section \ref{sec:scaling_impact}}\label{sm:simulation}

\begin{table}[H]
    \centering
    \begin{tabular}{lrrr}
\toprule
 & True $\varphi$ & $\varphi$ coverage & $T$ coverage \\
        \midrule
    Expectation & 0.20 & 0.21 & 0.83 \\ 
Geometric mean & 0.20 & 0.20 & 0.82 \\ 
No scaling & 0.20 & 0.21 & 0.83 \\ 
Expectation & 0.50 & 0.48 & 0.74 \\ 
Geometric mean & 0.50 & 0.47 & 0.74 \\ 
No scaling & 0.50 & 0.45 & 0.74 \\ 
Expectation & 0.80 & 0.47 & 0.52 \\ 
Geometric mean & 0.80 & 0.46 & 0.52 \\ 
No scaling & 0.80 & 0.46 & 0.53 \\ 
        \bottomrule
       \end{tabular}
       \caption{90\% coverage level for $\varphi$ and $T$ for the IG priors specification}\label{tab:IG_coverage}
\end{table}

\begin{table}[H]
    \centering
    \begin{tabular}{lrrr}
\toprule
 & True $\varphi$ & $\varphi$ coverage & $T$ coverage \\
        \midrule
   Expectation & 0.20 & 0.90 & 0.89 \\ 
Geometric mean & 0.20 & 0.90 & 0.89 \\ 
No scaling & 0.20 & 0.83 & 0.86 \\ 
Expectation & 0.50 & 0.81 & 0.79 \\ 
Geometric mean & 0.50 & 0.80 & 0.79 \\ 
No scaling & 0.50 & 0.79 & 0.75 \\ 
Expectation & 0.80 & 0.40 & 0.45 \\ 
Geometric mean & 0.80 & 0.41 & 0.46 \\ 
No scaling & 0.80 & 0.36 & 0.40 \\ 
        \bottomrule
       \end{tabular}
       \caption{90\% coverage level for $\varphi$ and $T$ for the PC priors specification}\label{tab:PC_coverage}
\end{table}

\begin{table}[hbt!]
    \centering
    \begin{tabular}{lrrr}
\toprule
 & True $\varphi$ & $\varphi$ coverage & $T$ coverage \\
        \midrule
    Expectation & 0.20 & 0.97 & 0.90 \\ 
Geometric mean & 0.20 & 0.97 & 0.89 \\ 
No scaling & 0.20 & 0.83 & 0.80 \\ 
Expectation & 0.50 & 0.98 & 0.93 \\ 
Geometric mean & 0.50 & 0.97 & 0.93 \\ 
No scaling & 0.50 & 0.93 & 0.94 \\ 
Expectation & 0.80 & 0.81 & 0.89 \\ 
Geometric mean & 0.80 & 0.82 & 0.89 \\ 
No scaling & 0.80 & 0.91 & 0.86 \\ 
        \bottomrule
       \end{tabular}
        \caption{90\% coverage level for $\varphi$ and $T$ for the VP prior specification}\label{tab:VP_coverage}
\end{table}

\newpage
\section{Additional simulation results from Section \ref{sec:Qmod_impact}}

\begin{figure}[hbt!]
    \centering
\includegraphics[width=\textwidth]{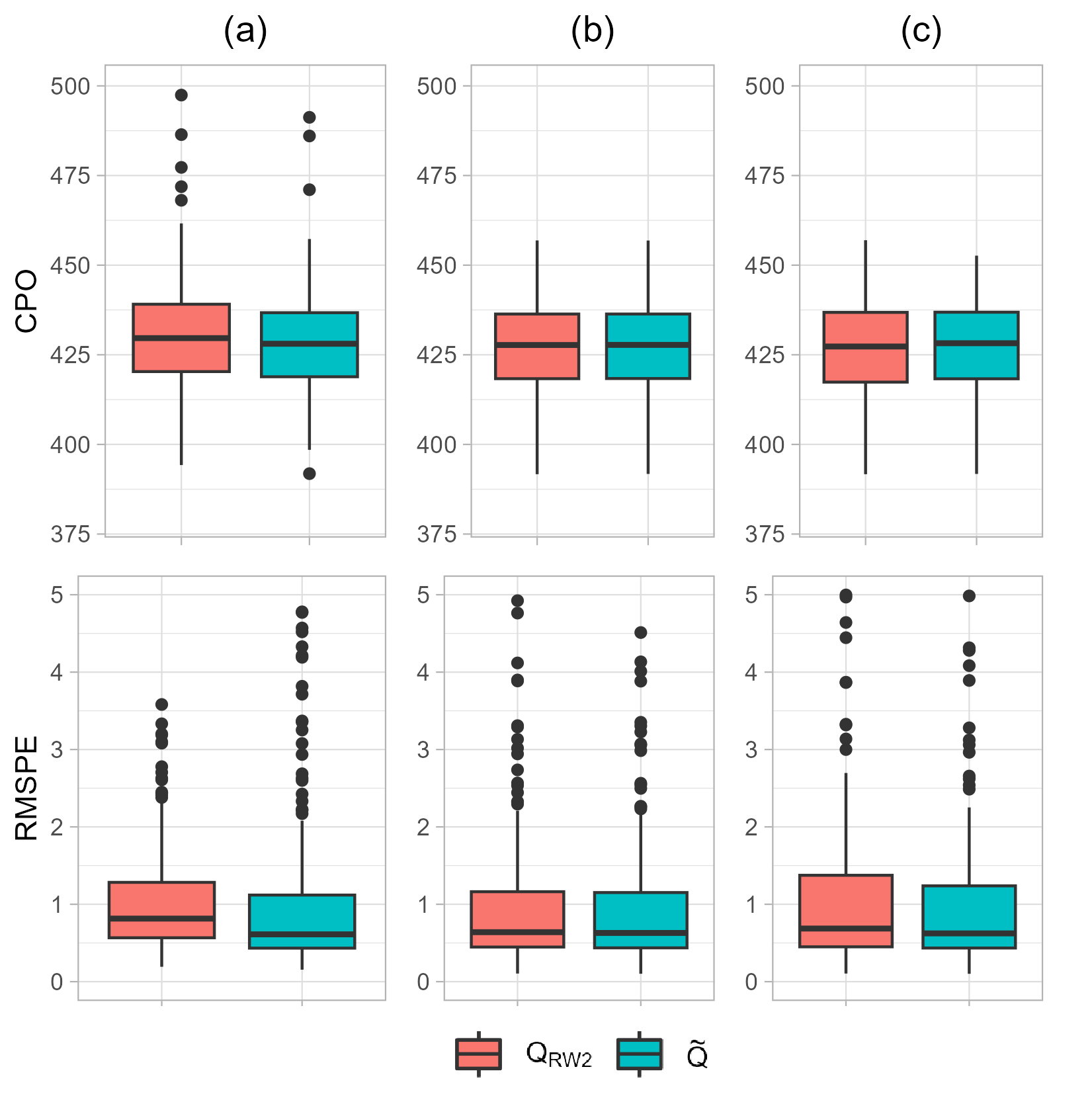}
   \caption{Negative sum of log conditional predictive ordinates (CPO) and Root Mean Squared Percentage Error (RMSPE) under the choice of the original $\boldsymbol{Q}_{\text{RW2}}$ or its modified version $\widetilde{\boldsymbol{Q}}$: (a) IG priors; (b) PC priors; (c) VP prior.}
\end{figure}

\end{document}